*Article*

# Dark Energy and Inflation from Gravitational Waves

**Leonid Marochnik**

Department of Physics, East-West Space Science Center, University of Maryland, College Park, MD 20742, USA; lmarochnik@gmail.com

**Abstract:** In this seven-part paper, we show that gravitational waves (classical and quantum) produce the accelerated de Sitter expansion at the start and at the end of the cosmological evolution of the Universe. In these periods, the Universe contains no matter fields but contains classical and quantum metric fluctuations, i.e., it is filled with classical and quantum gravitational waves. In such evolution of the Universe, dominated by gravitational waves, the de Sitter state is the exact solution to the self-consistent equations for classical and quantum gravitational waves and background geometry for the empty space-time with FLRW metric. In both classical and quantum cases, this solution is of the instanton origin since it is obtained in the Euclidean space of imaginary time with the subsequent analytic continuation to real time. The cosmological acceleration from gravitational waves provides a transparent physical explanation to the coincidence, threshold and "old cosmological constant" paradoxes of dark energy avoiding recourse to the anthropic principle. The cosmological acceleration from virtual gravitons at the start of the Universe evolution produces inflation, which is consistent with the observational data on CMB anisotropy. Section 1 is devoted to cosmological acceleration from classical gravitational waves. Section 2 is devoted to the theory of virtual gravitons in the Universe. Section 3 is devoted to cosmological acceleration from virtual gravitons. Section 4 discusses the consistency of the theory with observational data on dark energy and inflation. The discussion of mechanism of acceleration and cosmological scenario are contained in Sections 5 and 6. Appendix contains the theory of stochastic nonlinear gravitational waves of arbitrary wavelength and amplitude in an isotropic Universe.

**1. Cosmological Acceleration from Classical Gravitational Waves**

In Appendix, we obtained a set of exact self-consistent equations describing an ensemble of stochastic nonlinear classical gravitational waves of arbitrary amplitudes and wavelengths isotropic on average. In this paper, we use an approximation, in which these equations describe the backreaction of linear waves that affects the background metric, which in turn affects the state of gravitational waves. This also means that the interaction of gravitational waves is taken into account only through self-consistent background gravitational fields. In this approximation, superhorizon waves form the de Sitter state in the empty (with no matter fields) space-time with FLRW metric. This solution shows that the contemporary Universe must contain of the order of $10^{123}$ gravitons to create the observed Hubble constant. This number has nothing to do with vacuum energy, which is a possible solution to the "old cosmology constant problem".

*1.1. Introduction*

In the period of 2008–2017, we published a series of papers [1–7] on the cosmological acceleration generated by classical and quantum gravitational waves of the super-horizon wavelengths. In the present seven-part paper, these results are collected together with new results and are presented in the form of a complete theory of cosmological acceleration from gravitational waves. As such, it leads to a certain overlap with our previously published works. In these works, we showed that both classical gravitational waves and quantum gravitons of super-horizon wavelengths should lead to an accelerated expansion of homogeneous and isotropic empty (without matter) space-time in accordance with the de Sitter law. This is due to the fact that in the empty space-time there are always classical and quantum metric fluctuations, i.e., classical gravitational waves and gravitons. Thus, empty (or almost empty) space-time is always filled with gravitational waves and gravitons, which affect the background metric, which leads to its accelerated expansion. Although a few people seemed to doubt the reality of the gravitational waves predicted by the theory of relativity, the recent experimental discovery of gravitational waves [8] from merging black holes left apparently no doubt even among the skeptics. The modern Universe is emptying, ending its evolution. Today it is already 68% empty [9] and asymptotically will be completely empty since the density of matter in it is decreasing as $\rho_m \sim a(t)^{-3} \to 0$, $t \to \infty$ (if there are no unexpected discoveries showing that the Doomsday is being postponed). The early Universe, before the matter was born in it, was also presumed to be empty, so the effect of cosmological acceleration should work in both cases. In the case of very early Universe, there is no direct observational confirmation of cosmological acceleration but the hypothesis on inflation is very attractive due to its ability to solve three known major cosmological paradoxes (flatness, horizon and monopoles) [10–12]. In case of the contemporary Universe, the cosmological acceleration is the established observational fact, which is known as dark energy effect [13,14]. A common feature of both dark energy and inflation is exponentially rapid expansion of the Universe, which is generally accepted opinion. Another common feature is the fact in case of dark energy and hypothesis in case of inflation that both effects occur in an empty (or nearly empty) Universe. At the first time, the crucial importance of the emptiness of the spacetime was mentioned in [1]. Therefore, we believe that the effect of accelerating the expansion of empty space-time, discovered by us under the influence of classical and quantum gravitational waves, can serve as an explanation of these effects. In the case of classical gravitational waves, the de Sitter state of an empty homogeneous and isotropic space-time is the exact solution of self-consistent equations describing the backreaction effect of gravitational waves on the background geometry and the effect of this geometry on the behavior of gravitational waves [2]. In the case of gravitons, we constructed a one-loop finite self-consistent theory of the backreaction of gravitons on the background geometry and vice versa [3]. Same as the classical case, the de Sitter state is an exact solution of the self-consistent equations of one-loop quantum gravity [4]. In both the quantum and classical cases, a temporary transition to the Euclidean space of imaginary time is inevitable, followed by a return to real time [5]. In both cases, we are forced to treat time as a complex variable to replace integration along the real time axis by integrating along the imaginary axis. Therefore, the de Sitter solutions obtained are instanton solutions, if one uses generally accepted terminology. Is it possible to obtain a de Sitter solution (an exponentially rapid expansion of space-time), avoiding the transition to imaginary time? For classical gravitational waves, we have an unambiguous answer "no", as follows from the first part of this work devoted to classical gravitational waves. In the case of gravitons, the answer to this question is not so unambiguous. There are many papers in which the problem was posed of the backreaction of gravitons on the background geometry in the isotropic and homogeneous on the average space-time in the framework of the one-loop approximation of quantum gravity. Examples of such works can be found in following references [15–20]. Solving the same problem, the authors of all these papers obtained inconsistent results, as was mentioned by the authors of one of these papers [20]. This "irreproducible physics" arises because of the use of different methods for renormalization of divergences by different authors. Obviously, the result of solving a certain physical problem cannot depend on the calculation method. In addition, the renormalization procedure changes the original definition of the graviton (see below). The appearance of all of these works was caused by the following reason. One of the main problems of quantum gravity that distinguishes it from standard quantum field theories is a problem of the existence (or non-existence) of ghost-free gauges. Such gauges are unknown. DeWitt e.g. assumed that such gauges do exist [21]. In the frame of this assumption, he discussed general features of a path integral and basic equation of quantum geometrodynamics (Wheeler-DeWitt equation). Later this assumption was used in almost all works on the quantum theory of gravitons, becoming thus mainstream, and the results obtained in these works came to be regarded as a "standard". Typical examples of such works are mentioned above [15–20] (actual list of such works is much greater). In fact, the quantum theory of gravitons using this assumption does not look promising and here is why. There are at least two reasons that indicate that this is a dead-end branch. All these works (with no exception) use the linear parameterization of metric fluctuations of arbitrary wavelengths leading to the non-conservative energy-momentum tensor (EMT) and non-self-consistent set of quantum and classical equations. Existence of this problem is exhaustively documented in [22]. The correct parameterization leading to the conservative EMT and self-consistent set of quantum and classic equations was presented in our work [3] (see also Section 1.2 of the present paper, Appendix and Section 2). However, the most important problem is the appearance of divergences in the process of calculations. All these works lose any meanings after renormalization of original gravitational Lagrangian by quadratic counterterms. This procedure modifies the original Lagrangian and modifies in turn the original definition of the graviton, so that the theory starts with one definition of the graviton and finishes with another. The mathematical proof of this fact is given in Section X of [3]. In our theory of gravitons, we proceed from the assumption that the ghost-free gauges are unknown because they simply do not exist. In such a case, the appearance of ghosts in the path integral is inevitable with all the ensuing consequences [7]. The equations of one-loop approximation of quantum gravity (under the assumption that the ghost-free gauges do not exist) are derived in our work [7]. In Section 2 of this paper, this problem is discussed in sufficient detail. In the one-loop approximation, the quantum gravity of empty space-time must be finite on the mass shell [23]. We have shown that it can be made finite also off the mass shell [3]. In the framework of the finite one-loop approximation of quantum theory of gravitation, De Sitter's solution was obtained in both real time [4] and imaginary time (instanton solution) [2]. Interpretation of this solution in real time touches upon fundamental questions of theory of quantum gravity. As of today, such a theory does not yet exist, so the question of interpretation of this solution remains open. As to the instanton de Sitter solution, it is similar to the case of classical gravitational waves and its clear interpretation.

The outline of Section 1 is as follows. The set of exact equations describing the ensemble of stochastic nonlinear classical gravitational waves of arbitrary amplitudes and wavelengths isotropic on average is derived in Appendix. In quasi-linear approximation, the de Sitter solution for the empty space-time with the FLRW metric is obtained under affect of classical gravitational waves by two independent methods (Sections 1.2 and 1.3). In Section 1.4, we estimate the number of gravitons in the contemporary Universe based on the solution obtained.

*1.2. De Sitter Acceleration from Classical Gravitational Waves*

Isaacson [24] studied the back reaction of short classical gravitational waves on the background. The model of the empty Universe consisting of short classical gravitational waves was described for the first time in [25]. The energy momentum tensor (EMT) for super-long classical gravitational waves was constructed in [22]. The backreaction of super-long gravitational waves was studied in several works (see [22, 26] and references therein). The self-consistent back reaction of classical super-horizon wavelengths on the background metric and vice versa is considered in this paper. We show that a self-consistent solution to the back reaction problem for such waves is a de Sitter accelerated expansion[1].

The self-consistent back reaction of virtual gravitons of super-horizon wavelengths on the background metric and vice versa was considered in works [3, 4]. In the frame of one-loop finite quantum gravity it was shown that gravitons can form a quantum coherent condensate over the horizon scale of the Universe which provides the cosmological acceleration of the empty Universe described by the de Sitter law. It may give the impression that the de Sitter expansion of the empty Universe is a pure quantum effect. In fact, the de Sitter expansion of the empty Universe under backreaction of metric fluctuations (quantum or classical) takes place due to the fact of invariance of de Sitter state with respect to Wick rotation (see below).

---

[1] We already mentioned such a possibility in our work [9]. In this paper, we present a full consideration of the effect.

In the frame of the self-consistent approach, the state of classical gravitational waves (CGW) is determined by their interaction with background geometry, and the background geometry, in turn, depends on the state of CGW. In all works (with no exceptions) on the back reaction of the long gravitational waves the authors used linear parameterization of the metric tensor $g^{ik} = g_0^{ik} + h^{ik}$, where $g_0^{ik}$ and $h^{ik}$ are background metric tensor and its perturbation, respectively. This parameterization leads to the non-conservative energy momentum tensor for CGW and the non-self-consistent system of equations for the background metric and gravitational waves (see Appendix). The existence of this problem was exhaustively documented in [22]. In case of virtual gravitons, the solution to this problem was given in our works [3,4] where it was shown that it is necessary to work with the exponential parameterization (instead of linear one) corresponding to the use of normal coordinates in the functional space. The normal coordinates are introduced by exponential parameterization of the density of metric tensor [27].

$$(-g)^{1/2} g^{ik} = (-\overline{g})^{1/2} \overline{g}^{il} (\exp \Psi)_l^k = (-\overline{g})^{1/2} \overline{g}^{il} (\delta_l^k + \Psi_l^k + \frac{1}{2} \Psi_l^m \Psi_m^k + ...)$$

where $\overline{g}^{il}$, $i,l = 0,1,2,3$ is Minkowski metric and $\Psi_l^k$ is a tensor in the Minkowski space. Einstein's equations in normal coordinates are given in [7] and Section 2. In case of classical gravitational waves, the solution to this problem is given in the Appendix. With this parameterization, the energy-momentum tensor (EMT) of gravitational waves satisfies the conservative condition automatically. The Equations (A27)–(A31) of Appendix can be rewritten as (1)–(5) after Fourier transformation. They read

$$3H^2 = \kappa \rho_g \tag{1}$$

$$2\dot{H} + 3H^2 = -\kappa p_g \tag{2}$$

$$\rho_g = \frac{1}{8\kappa} \sum_{\mathbf{k}\sigma} < \dot{\psi}^*_{\mathbf{k}\sigma} \dot{\psi}_{\mathbf{k}\sigma} + \frac{k^2}{a^2} \psi^*_{\mathbf{k}\sigma} \psi_{\mathbf{k}\sigma} > \tag{3}$$

$$p_g = \frac{1}{8\kappa} \sum_{\mathbf{k}\sigma} < \dot{\psi}^*_{\mathbf{k}\sigma} \dot{\psi}_{\mathbf{k}\sigma} - \frac{k^2}{3a^2} \psi^*_{\mathbf{k}\sigma} \psi_{\mathbf{k}\sigma} > \tag{4}$$

where (1) and (2) are Einstein's equations for the background and (3) and (4) form the EMT of CGW. As is mentioned in the Appendix, in this approximation the dynamics of gravitational waves are described by Equation (5) in the linear approximation but the backreaction of gravitational waves on the background metric is described by Equations (1)–(4) in which the second order terms are retained. This means that the backreaction of waves affect the background metric, which in turn affects the state of gravitational waves. This also means that the interaction of gravitational waves is taken into account only through the self-consistent background gravitational field. In these equations $H = \dot{a}/a$ is the Hubble function; $\kappa = 8\pi G$; speed of light $c = 1$; $\rho_g$ and $p_g$ are the energy density and pressure of CGW; $\sigma$ is the polarization index; $\psi_{\mathbf{k}\sigma}$ is the Fourier images of tensor fluctuations (GW) that are gauge–invariant by definition; dots above functions denote derivatives over the physical time $t$; the superscript ($*$) is the sign of complex conjugation, and $a$ is the scale factor. The mathematically rigorous method of separating the background and the GW, which ensures the existence of the EMT, is based on averaging over polarizations $\sigma$ of CGW, and $<\psi_\alpha^\beta> \equiv 0$ if all polarizations are equivalent in the homogeneous isotropic GW ensemble [3]. About the gage invariance problem and the elimination of 3-scalar and 3-vector modes (see Section III.A of our work [3]). AS is well known, the equation for 3-tensor GW is

$$\psi_\alpha^\beta(t,\mathbf{x}) = \sum_{\mathbf{k}\sigma} Q_\alpha^\beta(\mathbf{k}\sigma)\psi_{\mathbf{k}\sigma}(t)e^{i\mathbf{k}\mathbf{x}}, \qquad \ddot\psi_{\mathbf{k}\sigma} + 3H\dot\psi_{\mathbf{k}\sigma} + \frac{k^2}{a^2}\psi_{\mathbf{k}\sigma} = 0 \qquad (5)$$

The transition from summation to integration in (3) and (4), taking into account the isotropy of space can be done in the following way

$$\sum_{\mathbf{k}}... \to \int d^3k/(2\pi)^3... = \int_0^\infty k^2 dk/2\pi^2.... \qquad (6)$$

It is also convenient to make the transition from physical time $t$ to the cosmological time $\eta = \int dt/a$. In such terms, Equations (3) and (4) can be rewritten in the following form

$$3\frac{a'^2}{a^4} = \kappa\rho_g = \frac{1}{16\pi^2}\int_0^\infty \frac{k^2}{a^2}dk(\sum_\sigma <\hat\psi'^*_{\mathbf{k}\sigma}\hat\psi'_{\mathbf{k}\sigma} + k^2\hat\psi^*_{\mathbf{k}\sigma}\hat\psi_{\mathbf{k}\sigma}>) \qquad (7)$$

$$2\frac{a''}{a^3} - \frac{a'^2}{a^4} = -\kappa p_g = -\frac{1}{16\pi^2}\int_0^\infty \frac{k^2}{a^2}dk(\sum_\sigma <\hat\psi'^*_{\mathbf{k}\sigma}\hat\psi'_{\mathbf{k}\sigma} - \frac{k^2}{3}\hat\psi^*_{\mathbf{k}\sigma}\hat\psi_{\mathbf{k}\sigma}>) \qquad (8)$$

Equation (5) in conformal time reads

$$\phi''_{\vec{k},\sigma} + (k^2 - \frac{a''}{a})\phi_{\vec{k},\sigma} = 0 \quad \psi_{\mathbf{k}\sigma} = \frac{1}{a}\phi_{\mathbf{k}\sigma} \qquad (9)$$

where primes are derivatives over cosmological time $\eta$. The de Sitter regime in terms of conformal time reads

$$a = -(H\eta)^{-1} \qquad (10)$$

The solution to (9) over the de Sitter background (10) reads

$$\psi_{\mathbf{k}\sigma}(\eta) = \frac{1}{a\sqrt{k}}[Q_{\mathbf{k}\sigma}(1-\frac{1}{ik\eta})e^{-ik\eta} + Q^*_{\mathbf{k}\sigma}(1+\frac{1}{ik\eta})e^{ik\eta}] \qquad (11)$$

Here $Q_{\mathbf{k}\sigma}$ is the integration constant. The solution (11) can be rewritten also in the following useful form

$$\psi_{\mathbf{k}\sigma}(x) = H \cdot k^{-3/2}[A_{\mathbf{k}\sigma} \cdot (x\cos x - \sin x) + B_{\mathbf{k}\sigma} \cdot (x\sin x + \cos x)], \quad x = k\eta \qquad (12)$$

Here $A_{\mathbf{k}\sigma} = 2\operatorname{Re} Q_{\mathbf{k}\sigma}$ and $B_{\mathbf{k}\sigma} = 2\operatorname{Im} Q_{\mathbf{k}\sigma}$.

To make sure that (10) is an exact solution to the set of Equations (7)–(9), we have to substitute (10) and (11) into (7) and (8) and find $H$ for which this set of equations is satisfied.

Substitution (11) into Equations (7) and (8) leads to divergent integrals.

The Euclidean space of imaginary time is defined by the metric

$$ds^2 = -d\tau^2 - a^2(\tau)(dx_1^2 + dx_2^2 + dx_3^2)$$

In this metric, Einstein's Equations (1)–(5) read

$$3H^2 = \kappa\rho_g$$

$$2\dot H + 3H^2 = -\kappa p_g$$

$$\rho_g = \frac{1}{8\kappa} \sum_{\mathbf{k}\sigma} < -\dot{\psi}^*_{\mathbf{k}\sigma}\dot{\psi}_{\mathbf{k}\sigma} + \frac{k^2}{a^2}\psi^*_{\mathbf{k}\sigma}\psi_{\mathbf{k}\sigma} >$$

$$p_g = \frac{1}{8\kappa} \sum_{\mathbf{k}\sigma} < -\dot{\psi}^*_{\mathbf{k}\sigma}\dot{\psi}_{\mathbf{k}\sigma} - \frac{k^2}{3a^2}\psi^*_{\mathbf{k}\sigma}\psi_{\mathbf{k}\sigma} >$$

$$\ddot{\psi}_{\mathbf{k}\sigma} + 3H\dot{\psi}_{\mathbf{k}\sigma} - \frac{k^2}{a^2}\psi_{\mathbf{k}\sigma} = 0$$

where dots indicate derivatives over $\tau = -it$.

After the transition from summation to integration in accordance with Equation (6), we get

$$\rho = \frac{1}{16\pi^2 a^2}\int_0^\infty k^2 dk \sum_\sigma < -\psi'_{k\sigma}\psi'^*_{k\sigma} + k^2\psi_{k\sigma}\psi^*_{k\sigma} > \quad (13)$$

$$\phi''_k - (k^2 + \frac{a''}{a})\phi_k = 0 \quad (14)$$

We get Equations (13) and (14) if we make Wick rotation $\eta = -i\varsigma$, $\xi = k\varsigma$ in Equations (7) and (8). Primes in Equations (13) and (14) indicate derivatives over $\varsigma$. De Sitter regime in Euclidean space reads

$$a = \frac{1}{iH\varsigma} \text{ or } a = \frac{1}{H_\tau\varsigma}, \quad H_\tau = \frac{1}{a}\frac{da}{d\tau} \quad (15)$$

Instead of (11), we get the following solution for (14) over the background

$$\psi_{k\sigma}(\xi) = -iH \cdot k^{-3/2}[b_{k\sigma}(\xi+1)e^{-\xi} + a^*_{k\sigma}(\xi-1)e^{\xi}] \quad (16)$$

To get a finite solution, one has to choose $a_{k\sigma} = 0$. Thus, we get

$$\psi_{k\sigma}(\xi) = -iH \cdot k^{-3/2} \cdot b_{k\sigma} \cdot (\xi+1)e^{-\xi} \quad (17)$$

Substitution (17) into (13) leads to the following equation for the energy density

$$\rho = \frac{H^4}{16\pi^2}\int_0^\infty \sum_\sigma <|b_{k\sigma}|^2>[\xi^2-(1+\xi)^2]e^{-2\xi}\xi d\xi \quad (18)$$

To simplify the analysis, we assume that $\sum_\sigma <|b_{k\sigma}|^2> = <|b_k|^2> = <|b|^2> = const$. $<|b|^2>$ is the mean square of the amplitude of the gravitational waves in the ensemble in imaginary time. The averaging is performed over the random but isotropic on average polarizations. We also assume that the spectrum is flat, i.e., it does not depend on wavelength $k$. The integral in Equation (18) reads

$$\int_0^\infty [\xi^2 - (1+\xi)^2]e^{-2\xi}\xi d\xi = -3/4$$

From (18) we get

$$\rho = -\frac{3<|b|^2>H^4}{64\pi^2} \quad (19)$$

In accordance with (15), after Wick rotation the LHS of (8) is $-3H^2$. From (8), (15) and (19) we get the following equation for the Hubble constant $H$

$$-3H^2 = -\frac{3\kappa <|b|^2> H^4}{64\pi^2} \tag{20}$$

Equation (9) is a Schrödinger-like equation with the "one-dimensional potential" $a''/a$. The only difference is that the spatial coordinate in the Schrödinger equation is changed for the time variable $\eta$ in Equation (7). In such terms, the solution (12) is a superposition of incident and reflected waves with respect to the "barrier" $x = \xi = 0$, and solution (17) is a transition wave. To make the analytic continuation from (17) back to the real time (reverse Wick rotation), we have to satisfy two boundary conditions (see also [2])[2]

$$\psi_{k\sigma}(\xi=0) = \psi_{k\sigma}(x=0) \tag{21}$$

$$\psi'_{k\sigma}(\xi=0) = \psi'_{k\sigma}(x=0) \tag{22}$$

The condition in Equation (22) is satisfied automatically because of $\psi'_{k\sigma}(\xi=0)=0$ and $\psi'_{k\sigma}(x=0)=0$ each separately. The condition (21) is satisfied if $b_{k\sigma} = iB_{k\sigma}$, which leads to the following

$$<|b|^2> = <|B|^2>$$

where $<|B|^2>$ is mean square of the amplitude of the gravitational waves in the ensemble in real time. The averaging is performed over the random but isotropic on average polarizations. We also assume that the spectrum is flat, i.e., it does not depend on wavelength $k$.

Finally, Equation (20) can be rewritten in the following form

$$3H^2 = \frac{3\kappa <|B|^2> H^4}{64\pi^2} \tag{23}$$

Thus, Equation (23) is a result of analytical continuation (19) and (20) to the Lorenzian space of real time[3]. It describes the de Sitter solution that takes place in real time in the empty Universe. Note that (23) is also the exact (not an approximate) solution to the same problem for the ensemble of randomly distributed but homogenous and isotropic in the average scalar fields (Section 5.3). As it follows from (23)

$$H^2 = \frac{8\pi}{G<|B|^2>} \tag{24}$$

---

[2] Note that these boundary conditions are the same at the potential barrier for the Schrödinger equation.

[3] Note that taking into account nonlinear terms can only change the numerical factor in this equation, but not its functional form. It is because the energy density in the equations of state (19) and (25) always takes the following form $\rho = C \cdot <|B|^2> H^4$, where $|B|^2$ is the action and $C$ is some numerical factor. This is a consequence of dimensionality [8]. As it was first established by Starobinsky [28], quantum corrections to the Einstein equations due to conformal anomalies leads to the appearance of de Sitter state with the equation of state $\rho = C_1 \cdot \hbar H^4$ instead of a Bing Bang singularity. It was shown by Zeldovich [29] that $C_1 \leq 100$ because it is of the order of the number of all elementary particles. It was shown by us [4] that in the case of gravitons, $C_1 \approx N_g \approx 10^{123}$ where $N_g$ is the number of gravitons in the contemporary Universe. As we show in Section 3.2, the equation of state of gravitons $\rho_g = \gamma N_g \hbar H^4 / 3\pi^2$ where $\gamma \sim 1$ follows from simple qualitative considerations, which do not require discussion of nonlinear effects (as is known, quantum gravity cannot be renormalized in higher loops [30]).

As was shown in [11], this de Sitter solution is formed by the waves with the wavelengths of the order of the horizon of events $\lambda \sim H^{-1}$. Note that $<B^2>$ is the action of the ensemble of randomly distributed gravitational waves that are isotropic on the average (averaging over their polarization). In Equations (23) and (24), $<B^2>^{1/2}$ is the root mean square of such an action.

Thus, a chaotic ensemble of classical gravitational waves which is homogenous and isotropic on the average generates the de Sitter expansion of the empty space. The speed of such expansion is

$$-p = \rho = \frac{3<|B|^2>H^4}{64\pi^2} \tag{25}$$

As was mentioned in the Footnote 3, for the first time, the de Sitter equation of state for the early Universe (with $\rho \sim H^4$) was obtained by Starobinsky [28], which was based on the quantum conformal anomalies. The notable fact is that in our case the same equation of state was produced by classical gravitational waves. The physics of the acceleration mechanism by gravitational waves is described in Section 5.

In view of the importance of the fact that the empty space is in the de Sitter state under the influence of classical gravitational waves, we give an independent proof of this fact by the use of the Bogoliubov-Born-Green-Kirkwood-Yvon hierarchy (BBGKY chain).

*1.3. De Sitter State of Empty Space as the Exact Solution to BBGKY Chain*

In this section, we obtain the de Sitter solution for the empty FLRW space by means of the BBGKY chain. For the first time, the solution was obtained in [4] for virtual gravitons. To build the BBGKY chain, one needs to introduce the gravitational wave spectral function $W_{\mathbf{k}}$ and its moments $W_n$

$$W_{\mathbf{k}} = \sum_{\sigma} <\psi^+_{\mathbf{k}\sigma}\psi_{\mathbf{k}\sigma}>$$

$$W_n = \sum_{\mathbf{k}} \frac{k^{2n}}{a^{2n}}\left(\sum_{\sigma} <\psi^+_{\mathbf{k}\sigma}\psi_{\mathbf{k}\sigma}>\right) \qquad n = 0,1,2,...,\infty \tag{26}$$

At this point, one can see a significant difference between classical GW and gravitons. In the quantum case, instead of (26) we have the following definition of the moments of the spectral function of gravitons [4]

$$W_n = \sum_{\mathbf{k}} \frac{k^{2n}}{a^{2n}}\left(\sum_{\sigma} <\Psi_g|\hat{\psi}^+_{\mathbf{k}\sigma}\hat{\psi}_{\mathbf{k}\sigma}|\Psi_g> - 2<\Psi_{gh}|\hat{\theta}^+_{\mathbf{k}}\hat{\theta}_{\mathbf{k}}|\Psi_{gh}>\right) \qquad n=0,1,2,...,\infty \tag{27}$$

where $\Psi_g$ and $\Psi_{gh}$ are graviton and ghost quantum state vectors; $\hat{\psi}_{\mathbf{k}\sigma}$ and $\hat{\theta}_{\mathbf{k}}$ are gravitons and ghosts operators, respectively. The most significant fact is the presence of Faddeev-Popov ghosts in the RHS of this equation (second term). One can see that in distinction to classical GW where the moments are always positive in accordance with (26), in the quantum case this is not so (for more details see [2,4] and references therein).

The derivation of the BBGKY chain can be found in Section V of work [3]. It reads

$$\dot{D} + 6HD + 4\dot{W}_1 + 16HW_1 = 0$$

$$\dddot{W}_n + 3(2n+3)H\ddot{W}_n + 3\left[\left(4n^2+12n+6\right)H^2 + (2n+1)\dot{H}\right]\dot{W}_n + \qquad n=1,...,\infty \tag{28}$$
$$+2n\left[2\left(2n^2+9n+9\right)H^3 + 6(n+2)H\dot{H} + \ddot{H}\right]W_n + 4\dot{W}_{n+1} + 8(n+2)HW_{n+1} = 0$$

$$D = \ddot{W}_0 + 3H\dot{W}_0 \tag{29}$$

Equation (28) form the BBGKY chain. Each equation of this chain connects the neighboring moments. Equation (28) have to be solved jointly with the Einstein Equations (1) and (2). In terms of $D$ and $W_1$ the energy density and pressure of gravitons are

$$3H^2 = \kappa \varepsilon_g \equiv \frac{1}{16}D + \frac{1}{4}W_1, \qquad 2\dot{H} + 3H^2 = -\kappa p_g \equiv \frac{1}{16}D + \frac{1}{12}W_1, \tag{30}$$

Instead of the original self-consistent system of (1) and (2) we now get the new self-consistent system of equations consisting of Einstein Equation (30) and the BBGKY chain (27) and (28). The energy-momentum tensor can be reduced to the form derived in [22] by identity transformations. As was shown in [4], the de Sitter solution is one of the exact solutions to the equations of the BBGKY chain for the empty FLRW space. One can check by simple substitution that it reads

$$H^2 = \frac{1}{36}W_1 \quad \varepsilon_g = -p_g = \frac{1}{12\kappa}W_1 \quad D = -\frac{8}{3}W_1, \quad a = a_0 e^{Ht} \tag{31}$$

$$W_{n+1} = -\frac{n(2n+3)(n+3)}{2(n+2)}H^2 W_n, \qquad n \geq 1 \tag{32}$$

Zero moment $W_0$, which has an infrared logarithmic singularity, is not contained in the expressions for the physical quantities, and for that reason, is not calculated. In the equation for $W_0$, the functions are differentiated in the integrand and the derivatives are combined in accordance with the definition (29). At the last step, the integrals that are calculated, already possesses no singularities. It can be seen from (32) that the signs of the moments $W_{n+1}/W_n < 0$ alternate. This alternation means that even moments are negative and this is in contradiction with their definition (26) (in distinction to the quantum case). This means that in real time $t$ we get a non-physical solution. The Wick rotation $t \to i\tau$ (the transition to imaginary time) leads to the following changes

$$t = i\tau; \quad H = -iH_\tau; \quad H_\tau = \frac{1}{a}\frac{da}{d\tau}; D \to -D^\tau; D^\tau = \frac{d^2}{d\tau^2}W_0 + 3H_\tau \frac{d}{d\tau}W_0 \tag{33}$$

This procedure removes alternation of the moments, and the solution (21) and (22) in imaginary time reads

$$-H_\tau^2 = \frac{1}{36}W_1; \quad D_\tau = \frac{8}{3}W_1 \tag{34}$$

$$W_{n+1} = \frac{n(2n+3)(n+3)}{2(n+2)}H_\tau^2 W_n, \qquad n \geq 1; \tag{35}$$

$$a = a_0 e^{H_\tau \tau} \tag{36}$$

Thus, (34)–(36) is the exact de Sitter solution to the BBKGY chain (27)–(30) in imaginary time. Due to the invariance of the de Sitter solution with respect to Wick rotation, we have this solution in real time too

$$a = a_0 e^{H_\tau \tau} = a_0 e^{iH \cdot (-it)} = a_0 e^{H \cdot t} \tag{37}$$

Thus, we obtained the de Sitter solution for the empty FLRW space-time by two independent methods described in Sections 1.2 and 1.3.

*1.4. Classical Gravitational Waves vs. Quantum Gravitons*

In accordance with the Planck Collaboration Cosmological Parameters [9], the Hubble constant reads

$$H = (67.3 \pm 1) \text{ km/sec} \cdot \text{Mpc} \tag{38}$$

To get the contemporary Hubble constant (38) which is created by CGW, one can estimate the appropriate action of CGW using (24). Substitution (38) into (24) gives

$$<|B|^2> \approx 1.8 \cdot 10^{89} \frac{\text{m}^2 \times \text{kg}}{\text{sec}} \tag{39}$$

Using the quantum language, one can say that we have an ensemble of super-long CGW with frequencies $H$ and energy $\hbar H$. If so, the ratio $<B^2>/\hbar \approx N_g$ gives us the number of gravitons under the de Sitter horizon in the Universe. The calculation of $N_g$ produces a familiar number

$$N_g \approx 1.7 \cdot 10^{123} \tag{40}$$

It is consistent with the number of virtual gravitons under the de Sitter horizon in the Universe which was calculated in the frame of quantum field theory [4,5] and Section 3.2. As was mentioned in [5] (see also Section 4.1.3), it could be a possible interpretation of the ratio $\rho_{Planck}/\rho_{Dark\,energy} \approx 10^{123}$ (so called "old cosmological constant problem"). In other words, this number is not a ratio of theoretical value of lambda term to its observational value but it simply is the number of gravitons under the horizon of the contemporary Universe. The consistency of existing observational data with the classical gravitational wave theory presented in this paper is discussed in Section 4.

*1.5. Conclusion*

There is no self-consistent solution to the problem of backreaction of classical gravitational waves in the empty space in real time. The transition to the Euclidian space of imaginary time produces the self-consistent de Sitter solution in such a space. This de Sitter solution in imaginary time was analytically continued into the Lorentzian space of real time. Transition to the Euclidean space of imaginary time is a *mandatory* procedure to get the de Sitter expansion of the empty space in real time under the back reaction of classical gravitational waves. In Section 2 of this paper, we consider the quantum theory of gravitational waves (gravitons). We emphasize the important fact that cosmological acceleration from gravitational waves whose existence in empty space-time is no longer in doubt, does not require additional hypotheses (even the most plausible hypotheses remain still hypotheses) to explain the causes of the effects of dark energy and inflation.

**2. Gravitons in the Universe[4]**

---

[4] This Section 2 "Gravitons in the Universe" actually should have two co-authors, Grigory Vereshkov and the author. A significant part of this paper is the content of our unpublished joint paper of 2013. Grigory died in 2014 and I am finishing our joint work alone. The problem is that by the end of his life, our views on the issue under discussion differed. And although much of the calculations in this paper belong to him, I do not consider it possible to put his name on this paper because of the differences in our points of view. He tried to show in the above-mentioned paper that the problem of the "ghost materialization" will find its explanation in the future quantum gravity theory. Perhaps this is so. On my part, I gradually came to the conclusion that "the ghost materialization" does not occur at all in the case of instanton solutions obtained in the Euclidean space of imaginary time. In the absence of a theory of quantum gravity, only such solutions have a physical meaning in my opinion. This idea is at the heart of all three parts of this paper.

In this section, we consider the quantum theory of gravitational waves (gravitons), and its direct connection with the fundamental problems of quantum gravity. In the non-stationary Universe, due to the conformal non-invariance and the zero rest mass of gravitons and ghosts, asymptotic states are absent, and the vacuum is unstable, both in the graviton sector and in the ghost sector. The irremovable quantum effect of the spontaneous production of particles in an unstable vacuum makes it impossible to retain for the ghost fields only the status of auxiliary virtual fields, which they had in the theory of the S-matrix. Ghosts formally acquire the properties of a second physical subsystem, dynamically equal in rights with the subsystem of gravitons. The appearance of ghosts means that extrapolating the theory of the S-matrix to the Universe as a whole derives the theory from the physical domain in which it was originally formulated as Faddeev's path integral with a measure obtained by the "decomposition of unity". The one-loop finiteness of the theory applies off the mass shell; One-loop equations have exact solutions that describe the macroscopic effects of quantum gravity—the condensation of gravitons, ghosts and instantons on the horizon scale of a non-stationary Universe. The instanton solutions are not accompanied by ghost materialization.

*2.1. Introduction*

We start with a short description of this work. The exact equations of quantum gravity in the Heisenberg representation are derived from the Faddeev path integral in the Hamiltonian gauge. For a macroscopic quantum-gravitational system, these equations are transformed to equations of the quantum theory of gravitons in a curved space-time with self-consistent geometry. The path integral, operator equations in the Heisenberg representation and the equations of the self-consistent theory of gravitons are related to each other by identity transformations. These contain the fact that in all formulations of the theory there is a specific ghost sector in the form of a scalar Grassmann field with a minimal coupling and a negatively determined energy-momentum tensor. In the S-matrix theory and in equivalent problems solved in the Heisenberg representation, the asymptotic states of ghosts are given by vacuum states. In these problems, the polarization of the ghost vacuum compensates for the vacuum polarization of nonphysical inertia fields. The selection rule excluding from the space of asymptotic states the state of ghosts with nonzero occupation numbers is mathematically ensured by the very existence of asymptotic states and by the stability of the vacuum.

In the non-stationary universe, due to the conformal non-invariance and the zero rest mass of gravitons and ghosts, asymptotic states are absent, and the vacuum is unstable, both in the graviton sector and in the ghost sector. The irremovable quantum effect of the spontaneous production of particles in an unstable vacuum makes it impossible to retain for the ghost fields only the status of auxiliary virtual fields, which they had in the theory of the S-matrix. Ghosts formally acquire the properties of a second physical subsystem, dynamically equal in rights with the subsystem of gravitons. The materialization of ghosts, strictly speaking, means that extrapolating the theory of the S-matrix to the universe as a whole derives the theory from the physical domain in which it was originally formulated as Faddeev's path integral with a measure obtained by the "decomposition of unity". This circumstance, however, does not preclude a detailed study of the intrinsic properties of the extrapolated theory. It is established that in this theory the ghost sector is unambiguously fixed; the one-loop finiteness of the theory applies off the mass shell; One-loop equations have exact solutions that describe the macroscopic effects of quantum gravity—the condensation of gravitons, ghosts and instantons on the horizon scale of a non-stationary Universe. It will be shown in the third part of the paper that in distinction to graviton-ghost condensates, instanton condensates do not produce the ghost materialization effect. Using Wick rotation (exactly as in the classical case), we again obtain de Sitter expansion of the empty space-time although the nature of this effect differs from the classical case.

In this part, we discuss two interrelated problems of the quantum theory of gravitation, and the related problem of the accelerated expansion of the Universe by the action of gravitons. First, we show that the existing quantum gravity in the formulation of Faddeev-Popov formally mathematically allows extrapolation from the theory of the S-matrix to the Universe as a whole. Secondly, solutions of the equations of extrapolated theory that predict a new class of physical phenomena will be described. These macroscopic effects of quantum gravity occur in the evolutionary stages of the universe. In the third part of this paper, we show that gravitons as well as classical gravitational waves generate the de Sitter accelerated expansion of the empty space-time with FLRW metric although the nature of this acceleration differs from the classical case. We gave exact solutions of one-loop equations of quantum gravity describing macroscopic effects for the first time in [4]. A detailed analysis of these solutions is given in [3]. In [7], the procedure for the transition from the Faddeev path integral [31] to the equations of quantum gravity in the Heisenberg representation and the subsequent transformation of these equations to the system of equations for quantum fields in a curved space-time with self-consistent geometry are described in detail. In the present work, we tried to discuss all these issues once again, drawing attention to the direct connection of the discussed problems with the fundamental problems of quantum gravity. The existence of macroscopic effects of quantum gravity is predicted from the following general considerations. Quanta of the gravitational field-gravitons—are Bose particles interacting with each other and with a macroscopic gravitational field responsible for the expansion of the universe as a whole. Because of the conformal non-invariance and zero rest mass of gravitons, their vacuum is unstable, and there is no threshold for the processes of vacuum polarization and particle creation. The most unstable are the so-called quasi-resonant modes whose wavelengths are comparable with the distance to the horizon of events. This means that, under certain conditions, the horizon of events can perform the functions of a "pump generator" that continuously works from the birth of the universe to our days. Given the gravitons belonging to the class of Bose particles, it is natural to assume that in the macroscopic system of gravitons the condensation effect of quasi-resonant modes (which is inherently related to the macroscopic quantum effects of superfluidity and superconductivity) is possible. In a condensate state, the state vector of a macroscopic system of gravitons becomes a coherent superposition of vectors corresponding to different occupation numbers of gravitons with the same wavelength of the order of the distance to the event horizon. Such a condensate should manifest itself as Dark Energy—a gravitating unstructured medium that homogeneously and isotropically fills the entire space of the Universe. The consideration of macroscopic effects of quantum gravity is possible only if there is a theory capable to predict (quantitatively) these effects. Comparing the mathematical constructions of the quantum Yang-Mills theory (Section 2.2) and quantum gravity (Section 2.3), we show the non-alternating character of the operator equations for gravitons and ghosts in the Heisenberg representation derived from the Faddeev path integral [31] from the extended phase space of gravitons and ghosts in the Hamiltonian gauge. In Section 4, the exact equations of the theory of gravitons and ghosts in the curved space-time with classical self-consistent geometry are obtained by identity transformations of these equations. In Section 2.6, the self-consistent theory of gravitons and ghosts in an isotropic non-stationary Universe is presented. Macroscopic quantum gravity phenomena are described by exact solutions of one-loop equations of this theory.

The mathematical formalism of the theory unambiguously testifies that the Faddeev-Popov (FP) ghosts are an ineradicable element of the quantum theory of gravitation in any of its representations and formulations. This fact is a direct consequence of the impossibility of an unambiguous choice of the frame of reference in space-time with dynamic geometry. Four gauge conditions imposed on the dynamic metric fix a certain class of reference frames, but the equations of the theory remain degenerate with respect to residual transformations of the group of diffeomorphisms corresponding to transformations of reference frames inside the chosen class.

Fictitious inertia fields corresponding to residual transformations cannot be globally separated from the true gravitational field created by interacting particles. In the quantum theory, fictitious virtual inertia fields are mixed with true virtual gravitons and, thereby, generate a fictitious contribution to the polarization of the graviton vacuum. The idea of compensating for the contribution of fictitious inertial fields with the fictitious field of ghost belongs to Feynman [32]. In the theory of a graviton S-matrix based on the FP path integral [33], ghosts have the status of an auxiliary mathematical object used to define a measure of functional integration with respect to the space-time metric with incomplete removal of degeneracy. The question of the status of ghosts in the theory of macroscopic quantum-gravitational phenomena is much more nontrivial. It is necessary to pay attention to the following circumstances. First, the equations of quantum gravity in the Heisenberg representation exist only in the distinguished Hamiltonian gauge, which corresponds to a fixed ghost sector in the form of a Grassmann scalar field theory with a minimum coupling with gravity. Secondly, this field is transformed to an (almost) standard bosonic complex scalar field by separating the Grassmannian units from the ghost fields in the form of multiplicative factors.

The only difference from the standard scalar field is that the energy-momentum tensor of bosonized ghosts enters the Einstein equation with an "incorrect" sign. The third (most important and critical for the theory) circumstance is the fact that in a macroscopic quantum-gravitational system the vacuum of bosonized ghosts is unstable. In an unstable vacuum, ghosts cannot be in a state with zero occupation numbers. This means that ghosts can no longer be considered only as compensators for the contributions of nonphysical inertia fields to the polarization of a graviton vacuum. The inevitable changes in the state of ghosts during the evolution of the system forces us to seek a new interpretation of ghosts as the second physical subsystem dynamically equivalent with the subsystem of gravitons. The ghost materialization is a direct mathematical consequence of the formulation of the problem, which is the only possible within the framework of the existing quantum theory of gravity. Namely, the postulate that the concrete representation and adequate formalism of the theory of macroscopic quantum gravitational phenomena should be obtained by extrapolating the theory of the FP graviton S-matrix [33] to the Universe as a whole is uncontested.

This approach preserves the most important property of the theory—the one-loop finiteness of quantum gravity without matter fields. We emphasize that this theory has this property both on the mass shell of gravitons [23] and off the mass shell[5]. It is the fact of the finiteness of the theory off the mass surface that makes it possible to obtain exact solutions of the cosmological one-loop equations of quantum gravity that explicitly demonstrates the existence of quantum-gravitational Bose condensates possessing the properties of Dark Energy. The exact solutions of one-loop equations of instanton type do not accompanied by the ghost materialization and reason for that is discussed in Section 3.

## 2.2. The Scheme of the Yang-Mills Quantum Theory

The features of the problem of extrapolation can be demonstrated by comparing the theories of Einstein and Yang-Mills. In the quantum theory of Yang-Mills, there is no problem of extrapolation, since there exist a universal representation that makes it possible to set problems both in the theory of the S-matrix and in the theory of the macroscopic system of quanta of gauge fields. We have in mind the operator of the Yang-Mills equations in the Heisenberg representation in the ghost-free Hamiltonian gauge $A^{\alpha}{}_0 = 0$. These equations are written in the Hamiltonian form with canonical commutation relations for the operators of generalized coordinates and moments [34].

$$\partial_0 \hat{\pi}_a^{\alpha} = i[\hat{H}, \hat{\pi}_a^{\alpha}]_- , \quad \partial_0 \hat{A}_{\alpha}^a = i[\hat{H}, \hat{A}_{\alpha}^a]_- ,$$
$$[\hat{\pi}_a^{\alpha}(t,\vec{x}), \hat{A}_{\beta}^b(t,\vec{x})]_- = -i\delta_a^b \delta_{\beta}^{\alpha} \delta(\vec{x}-\vec{x}');$$
$$\hat{H} = \int \hat{\tilde{H}} d^3x \quad \hat{\tilde{H}} = -\hat{\pi}_a^{\alpha} \hat{\pi}_{\alpha}^a + \hat{F}_{\alpha\beta}^a \hat{F}_a^{\alpha\beta} \tag{41}$$

where $\hat{F}_a^{\alpha\beta}$ are the strengths of Yang-Mills field; $\hat{\tilde{H}}$ and $\hat{H}$ are the density of Hamiltonian and Hamiltonian; $a, b$ are indexes of inner symmetry; $\alpha, \beta = 1, 2, 3$ are metric indexes that belong to the Minkowski space (all operations with metric indexes are performed with the Minkowski metric).

---

[5] So far, for the description of gravitons in the non-stationary Universe, the models used in the literature have no mathematical connection with the theory of the graviton S-matrix at the level of identity transformations (see, e.g., [15–19]). We do not discuss these models, since they are not relevant to the formalism of the existing quantum gravity. We note only that in all the published papers on the theory of gravitons in the non-stationary Universe, the condition of one-loop finiteness of quantum gravity is not satisfied.

The transitions from (41) to the Schrödinger representation and the interaction representation are carried out by standard unitary transformations of operators and state vectors. In the construction of the S-matrix in the interaction representation, an additional axiom on the existence of asymptotic states of free particles is introduced into the theory. It is well-known that this additional axiom makes sense only in the perturbation theory, although the initial operator equations given in (41) in the Heisenberg representation can also be used outside perturbation theory without the assumption of the existence of asymptotic states.

Equations (41) (or the Schrödinger equation derived from (41)) specify the procedure for calculating the matrix element of the evolution operator. The path integral in the Yang-Mills theory is initially introduced as an integral over the canonical Hamiltonian variables in the Hamiltonian gauge:

$$<out|in> = \int \exp\{i \int [\hat{\pi}_a^\alpha \partial_0 \hat{A}_\alpha^a - \tilde{\tilde{H}}(\hat{\pi}_a^\alpha, \hat{A}_\alpha^a)] d^4x\} \prod_x \prod_{a,\alpha} d\hat{\pi}_a^\alpha d\hat{A}_\alpha^a \quad (42)$$

By definition, the integral (42) is a mathematically equivalent way of computing the matrix element of the evolution operator. The application of the path integral to the calculation of the S-matrix initiates its further transformations. The transition to the integral over Lagrangian variables in the Hamiltonian gauge, and then the transition to the integral over all Lagrangian variables with the introduction of the Hamiltonian gauge into the measure of integration have the status of identical transformations. Up to this stage inclusively, there are no ghosts in the quantum Yang-Mills theory.

The ghosts in the Yang-Mills theory appear as auxiliary mathematical objects at the stage of converting of the measure of integration to the gauges, under which the residual degeneracy of the equations of the theory is preserved. These transformations include operations of "unit expansions" and permutations of the order of integration in divergent integrals over a gauge group. Strictly speaking, the use of these operations involves the calculation of the functional integral by perturbation theory under asymptotic boundary conditions. The assignment of such conditions is mathematically equivalent to the introduction of an axiom on the existence of asymptotic states in the operator formalism. The S-matrix constructed on the basis of an arbitrarily gauged path integral, is defined by extending the definition of an obvious physical statement: the space of non-vacuum asymptotic states is not changed under transformations of the integration measure from the ghost-free Hamiltonian gauge to other gauges generating the ghost sector. The mathematical realization of this statement is the selection rules: the asymptotic states of gravitons with arbitrary occupation numbers and vacuum states of ghosts with zero occupation numbers are physical. These selection rules automatically ensure the gauge invariance of the S-matrix, and the introduction of these rules is possible insofar as the vacuum is stable in the theory of the perturbative S-matrix.

*2.3. Scheme of Quantum Theory of Gravitation*

The quantum theory of gravity cannot be constructed according to the scheme described above for the quantum theory of Yang-Mills fields. The reason has already been pointed out earlier. In the theory there are no ghost-free gauges that globally separate the inertia fields from the true gravitational field. Thus, there is no ghost-free operator Hamiltonian formalism, which could be used as an initial representation of the quantum theory of gravity[6]. The modern quantum theory of gravity, which allows for the quantitative analysis of physical processes, exists only in the form of the theory of a perturbative S-matrix based on a gauged FP path integral. The computational capabilities of the theory are limited either by the scattering matrix of any gravitationally interacting fields in the tree approximation, or by a purely graviton S-matrix in the one-loop approximation, in which quantum gravity without matter fields is finite[7]. The procedure for extrapolating the theory of the graviton S-matrix to the Universe as a whole consists of two stages. In the first stage, operator equations in the Hamiltonian form in the Heisenberg representation are derived. The solutions of these equations for the same boundary conditions and selection rules for which the path integral is calculated must automatically reproduce the S-matrix of the FP theory. It is obvious that the S-matrices coincide only if the operator equations in the Heisenberg representation with the canonical quantization of gravitons and ghosts contain exactly the same ghost sector as the path integral from which they are derived. The actual extrapolation is carried out at the second stage. At the level of the formal postulate, it is asserted that the operator equations in the Heisenberg representation can be used to construct the theory of a macroscopic quantum-gravitational system whose structure corresponds not to asymptotic but to periodic boundary conditions. Such a system is an infinite non-stationary universe.

---

[6] The Hamiltonian formalisms of Dirac [35] and Arnowitt-Deser-Misner [36] use gauges that generate ghosts. The ghosts sector in [35,36] is not taken into account. (These works were performed long before the role of ghosts in the restoration of unitarity of the S-matrix was clarified.)

The Einstein and Yang-Mills theories have a common feature: in both cases, the operator equations in the Heisenberg representation do not exist in all gauges, but only in the chosen ones. A gauge should reduce the number of dynamic variables so that variations in the remaining variables would give only the equations of motion, and the constraint equations would act as the first integrals of the motion of these equations. This property is possessed by the Hamiltonian gauge in the Yang-Mills theory. The classical theory of gravitation has a synchronous gauge in normal coordinates, which has exactly this property (see Section II.1 of work ([7]). The normal coordinates are introduced by exponential parameterization of the density of the metric tensor:

$$(-g)^{1/2} g^{ik} = (-\bar{g})^{1/2} \bar{g}^{il} (\exp \Psi)_l^k = (-\bar{g})^{1/2} \bar{g}^{il} (\delta_l^k + \Psi_l^{\ k} + \frac{1}{2} \Psi_l^{\ m} \Psi_m^{\ k} + ...) \qquad (43)$$

where $\bar{g}^{il}$, $i,l = 0,1,2,3$ is Minkowski metric and $\Psi_l^{\ k}$ is a tensor in the Minkowski space. The Hamiltonian gauge is given by the conditions $(-\hat{g})^{1/2} g^{00} = 1$, $(-\hat{g})^{1/2} g^{0\alpha} = 0$ from which follows $\hat{\Psi}_0^{\ i} = 0$. This gauge, however, does not completely separate the inertia fields from the true gravitational field.

The Einstein Equations in Normal Coordinates Can Be Obtained in the Following Way.

Upon obtaining the exact equations of the theory of gravity in the Hamilton form, the gravitational field can be regarded as the deviation of the metric from the metric of Minkowski space $\bar{g}_{ik} = \text{diag}(1,-1,-1,-1)$. Normal coordinates of the gravitational field of $\hat{\Psi}_i^k$ are given by the exponential parameterization of the density of the contravariant metric [37]

$$\sqrt{-\hat{g}}\hat{g}^{ik} = \sqrt{-\bar{g}}\bar{g}^{il}\,\hat{g}_l^k,$$
$$\hat{g}_l^k \equiv (\exp \hat{\Psi})_l^k = \delta_l^k + \hat{\Psi}_l^k + \frac{1}{2}\hat{\Psi}_l^m\hat{\Psi}_m^k + \cdots \qquad (44)$$

The density of the gravitational Lagrangian as a function of normal coordinates reads

$$\mathcal{L}_{grav} = -\frac{1}{2\varkappa}\sqrt{-\hat{g}}\hat{g}^{ik}\hat{R}_{ik} =$$
$$= \frac{1}{8\varkappa}\,\hat{g}_k^l\left(\hat{\Psi}_n^{m,k}\hat{\Psi}_{m,l}^n - \frac{1}{2}\hat{\Psi}^{,k}\hat{\Psi}_{,l} - 2\hat{\Psi}_n^{k,m}\hat{\Psi}_{m,l}^n\right) + \dots, \qquad (45)$$

where dots denote a full derivative which does not contribute to the equations of motion. A variation of the action of the gravity theory over the normal coordinates leads to the Einstein equation

---

[7] In higher orders of perturbation theory, quantum gravity is not renormalizable [30]. Multi-loop calculations can be carried out only in finite super-gravities.

$$\sqrt{-\hat{g}}\,\hat{g}^{kl}\hat{R}_{il} \equiv$$
$$\frac{1}{2}\left(\hat{g}^{ml}\widehat{\Psi}^{k}_{i,m} - \hat{g}^{km}\widehat{\Psi}^{l}_{i,m} - \hat{g}^{ml}\Psi^{k}_{m,i} - \frac{1}{2}\delta^{k}_{i}\,\hat{g}^{ml}\widehat{\Psi}_{,m}\right)_{,l} - $$
$$-\frac{1}{4}\,\hat{g}^{kl}\left(\widehat{\Psi}^{n}_{m,i}\widehat{\Psi}^{m}_{n,l} - \frac{1}{2}\widehat{\Psi}_{,i}\widehat{\Psi}_{,l} - 2\widehat{\Psi}^{n}_{l,m}\widehat{\Psi}^{m}_{n,i}\right) = 0 \,. \tag{46}$$

The same equations can be rewritten in the form

$$\hat{\mathcal{E}}^{k}_{i} \equiv \sqrt{-\hat{g}}\,\hat{g}^{kl}\hat{R}_{il} - \frac{1}{2}\delta^{k}_{i}\sqrt{-\hat{g}}\,\hat{g}^{ml}\hat{R}_{ml} = 0 \tag{47}$$

In accordance with the general properties of Einstein equations, $\sqrt{-\hat{g}}\,\hat{g}^{\beta l}\hat{R}_{\alpha l} = 0$ are equations of motion ($\alpha,\beta = 1,2,3$), and $\hat{\mathcal{E}}^{0}_{i} = 0$ are equations of constraints. Note also the following form in which the Bianchi identity can be presented:

$$\frac{\partial \hat{\mathcal{E}}^{k}_{i}}{\partial x^{k}} + \frac{1}{2}\frac{\partial \widehat{\Psi}^{l}_{k}}{\partial x^{i}}\left(\hat{\mathcal{E}}^{k}_{l} - \frac{1}{2}\delta^{k}_{l}\hat{\mathcal{E}}\right) \equiv 0 \tag{48}$$

In Section II.2 of [7], it is shown that a non-removable inertia field has one degree of freedom, and the corresponding parameter of residual infinitesimal transformations satisfies the wave equation for a scalar field with a minimum coupling with gravity. Using this result and taking into account the general properties of the functional integration formalism, one can immediately predict that in the effective Lagrangian of the quantum theory of gravitation in a Hamilton gauge there will be one complex Grassmannian scalar field of ghosts that satisfies the formally general covariant Klein-Gordon-Fock equation. The existence of a Hamiltonian formalism for a scalar field is obvious; therefore, it is immediately possible to write out Einstein's operator equations in the Hamiltonian gauge in the Heisenberg representation. On the right side of these equations are the energy-momentum tensor of the ghosts and the operators of the gravitational and ghost fields will satisfy, respectively, the commutation and anti-commutation relations. For these reasons, Einstein's operator equations in the Hamiltonian gauge were written out in [3, 4] as initial equations without detailed explanations.

An exhaustive mathematical proof of the existence of quantum gravity in the Heisenberg representation with the canonical rules for quantizing gravitons and ghosts is presented in [7]. The proof is based on the general theory of Hamiltonian systems with incomplete removal of degeneracy formulated by Faddeev in [38]. In the application to quantum gravity, this theory is formulated as a path integral over conjugate Hamiltonian variables [31]. (See Equations (5) and (6) in [39] or Equations (22) and (24) in [27]). Faddeev's path integral in the Hamilton gauge reads

$$< \text{out}|\text{in}> = $$
$$\int \exp\left\{i\int\left[\hat{\pi}^{\nu}_{\mu}\partial_{0}\widehat{\Psi}^{\mu}_{\nu} - \mathcal{H}_{grav}(\hat{\pi}^{\nu}_{\mu},\widehat{\Psi}^{\nu}_{\mu})\right]d^{4}x\right\} \times$$
$$\times (\det \widehat{M}^{i}_{k})\prod_{x}\prod_{\mu\leq\nu}d\hat{\pi}^{\nu}_{\mu}d\widehat{\Psi}^{\nu}_{\mu}\,, \tag{49}$$

$$\mathcal{H}_{grav} = 2\varkappa\left(\hat{\pi}^{\mu}_{\nu}\hat{\pi}^{\nu}_{\mu} - \hat{\pi}\hat{\pi}\right) -$$
$$-\frac{1}{8\varkappa}\,\hat{g}^{\sigma}_{\rho}\left(\widehat{\Psi}^{\mu,\rho}_{\nu}\widehat{\Psi}^{\nu}_{\mu,\sigma} - \frac{1}{2}\widehat{\Psi}^{,\rho}\widehat{\Psi}_{,\sigma} - 2\widehat{\Psi}^{\rho,\mu}_{\nu}\widehat{\Psi}^{\nu}_{\mu,\sigma}\right) \tag{50}$$

Equation (50) is the density of gravitational Hamiltonian

$$\widehat{M}^{i}_{k} = \begin{pmatrix} \partial_{0} & -\partial_{\alpha} \\ \sqrt{-\hat{g}}\,\hat{g}^{\alpha\beta}\partial_{\beta} & \delta^{\alpha}_{\beta}\partial_{0} \end{pmatrix} \tag{51}$$

Equation (51) is the operator for the equations of residual degeneracy corresponding to the Hamiltonian gauge. After that, $\det \widehat{M}_{k}^{\;i}$ is localized in terms of Hamilton variables for anti-commute ghosts fields

$$\det \hat{M}^i_{\ k} = \int \exp \left\{ i \int [\mathcal{P} \cdot \partial_0 \theta + \partial_0 \bar{\theta} \cdot \bar{\mathcal{P}} - \right.$$
$$\left. - \mathcal{H}_{ghost}(\mathcal{P}, \theta, \bar{\mathcal{P}}, \bar{\theta}, \hat{\Psi}^\nu_\mu)] d^4 x \right\} \prod_x d\bar{\theta} d\bar{\mathcal{P}} d\theta d\mathcal{P} \tag{52}$$

where

$$\mathcal{H}_{ghost}(\mathcal{P}, \theta, \bar{\mathcal{P}}, \bar{\theta}, \hat{\Psi}^\nu_\mu) = -8\varkappa \mathcal{P}\bar{\mathcal{P}} + \frac{1}{8\varkappa} \cdot \hat{g}^\nu_\mu \bar{\theta}_{,\nu} \theta^{,\mu} \tag{53}$$

is the density of the ghost Hamiltonian. Substitution of (50) into (51) produces the path integral over the extended space of gravitons and ghosts. The factorization of the measure of integration allows to immediately considering this integral as a matrix element of the operator of evolution, which is calculated over the basis vectors of gravitons and ghost. A mathematically equivalent method for computing this matrix element is based on solutions of operator equations in the Heisenberg representation formulated in the extended phase space and supplemented with the canonical rules for quantizing the gravitational and ghost fields.

$$\partial_0 \hat{\pi}_\beta^{\ \alpha} = i[H, \hat{\pi}_\beta^{\ \alpha}]_- \qquad \partial_0 \hat{\Psi}_\alpha^{\ \beta} = i[H, \hat{\Psi}_\alpha^{\ \beta}]_-$$
$$[\hat{\pi}_\mu^{\ \nu}(t, \vec{x}), \hat{\Psi}_\rho^{\ \sigma}(t, \vec{x})]_- = -i\delta_\mu^{\ \sigma} \delta_\rho^{\ \nu} \delta(\vec{x} - \vec{x}') \tag{54}$$

$$\partial_0 \mathcal{P} = i[H, \mathcal{P}]_- \qquad \partial_0 \theta = i[H, \theta]_-$$
$$\partial_0 \bar{\mathcal{P}} = i[H, \bar{\mathcal{P}}]_- \qquad \partial_0 \bar{\theta} = i[H, \bar{\theta}]_- \tag{55}$$
$$[\theta(t, \vec{x}), \mathcal{P}(t, \vec{x})]_+ = [\bar{\theta}(t, \vec{x}), \bar{\mathcal{P}}(t, \vec{x})]_+ = i\delta(\vec{x} - \vec{x}')$$

In (54) and (55), there is a complete Hamiltonian of gravitons and ghosts, which reads

$$H = \int (\mathcal{H}_{grav} + \mathcal{H}_{ghost}) d^3 x \tag{56}$$

In the frame of perturbation theory, the Hamiltonian (56) is represented in the form of sum of Hamiltonian of free fields and the Hamiltonian of interactions, so that $H = H_0 + H_{int}$. In the interaction representation, both Hamiltonians are functionals of the free field operators. It is easy to obtain the explicit forms for $H_0$ and $H_{int}$ from (50) and (53) by extraction of delta-symbol from the operator exponent

$$(\exp \hat{\Psi}_{(0)})_\rho^{\ \sigma} = \delta_\rho^{\ \sigma} + [(\exp \hat{\Psi}_{(0)})_\rho^{\ \sigma} - \delta_\rho^{\ \sigma}]$$

The equations for the operators of free fields are obtained from (54) and (55) by exchange $H$ with $H_0$. The scattering operator is represented by Hamiltonian of interactions $H_{int}$ as a standard chronological exponent.

From the theory in the representation of the interaction supplemented by the axiom of the existence of asymptotic states and the rules of selection of physical states, we obtain an S-matrix that is identical to the S-matrix obtained from the Faddeev path integral (49), or from an arbitrarily gauged FP' path integral [33] taking into account identical transformations using the "unity decomposition". This result clearly demonstrates the following key statement.

*The existing quantum theory of gravity is reduced by means of identical transformations to the operator equations in the Heisenberg representation in the Hamiltonian gauge with the canonical rules for quantizing gravitons and ghosts.*

We draw attention to the fact that the schemes of quantum theories of Einstein and Yang-Mills consist of the same elements, but are different in relation to each other.

*For Yang-Mills theory:* operator equations in the Heisenberg representation in a ghost-free Hamiltonian gauge (41)→ Faddeev's path integral in the Hamiltonian gauge without the ghost sector (42)→ FP path integral in an arbitrary gauge with the ghost sector.

*For Einstein theory*: FP's path integral in an arbitrary gauge with the ghost sector → Faddeev's path integral in a Hamiltonian gauge with a specific ghost sector in the form of a scalar wave field (49)→ operator equations in the Heisenberg representation in a Hamiltonian gauge with the same ghost sector (54) and (55).

Equations (54) and (55), together with the constraint equations (contained in these as the first integrals of motion) are the Einstein operator equations with the energy-momentum tensor of scalar ghosts and the operator equations for ghosts.

$$\sqrt{-\hat{g}}\hat{g}^{kl}\hat{R}_{il} - \frac{1}{2}\delta_i^k \sqrt{-\hat{g}}\hat{g}^{ml}\hat{R}_{ml} =$$
$$-\frac{1}{4}[\sqrt{-\hat{g}}\hat{g}^{kl}(\partial_l\bar{\theta}\cdot\partial_i\theta + \partial_i\bar{\theta}\cdot\partial_l\theta) - \delta_i^k\sqrt{-\hat{g}}\hat{g}^{lm}\partial_l\bar{\theta}\cdot\partial_m\theta] \quad (57)$$

$$\partial_i\sqrt{-\hat{g}}\hat{g}^{ik}\partial_k\theta = 0, \qquad \partial_i\sqrt{-\hat{g}}\hat{g}^{ik}\partial_k\bar{\theta} = 0 \quad (58)$$

There are two arguments that allow us to assert that Equations (57) and (58) are the only possible equations of quantum gravity in the Heisenberg representation. First, we should pay attention to the uniqueness of the path integral of Faddeev [31] over the canonically conjugate Hamiltonian variables. Second, the Hamiltonian calibration is unambiguously fixed. It provides for the hamiltonization of the ghost sector. It leads to the path integral with the standard definition of the matrix element of the evolution operator, and, finally, it preserves the constraint equations in the form of the first integrals of the equations of motion.

Equations (57) and (58) differ from the theory of a scalar field in curved space-time only in the fact that the scalar field obeys an anti-commutative Grassmann algebra and, as a consequence, is quantized by anti-commutation relations. In the representation of interaction at the Feynman diagram level, the Grassmann algebra provides the "work" of the ghosts as compensators for the contribution of fictitious inertia fields to the polarization of the vacuum. In the case where the equivalent problem is solved in the Heisenberg representation, the ghost field is automatically subjected to the bosonization operation. This operation reduces to the allocation of Grassmann units in the form of multiplicative factors in front of operators of a complex field with a commutative algebra.

$$\theta = u\hat{\varphi}, \quad \bar{\theta} = \bar{u}\varphi^+, \quad \bar{u}u = -u\bar{u} = 1 \quad (59)$$

After this procedure, the anti-commutation relations for the quantum Grassmann field become standard commutation relations for the quantum complex scalar field. The origin of this field from the FP ghosts is manifested only in the fact that its energy-momentum tensor appears on the right side of Einstein's operator equations with an "incorrect" (anti-gravitating) sign. In other words, we get again (57) and (58) but with following change $\bar{\theta} \to \hat{\varphi}^+$; $\theta \to \hat{\varphi}$.

$$(-\hat{g})^{1/2}\hat{g}^{kl}\hat{R}_{il} - \frac{1}{2}\delta_i^k(-\hat{g})^{1/2}\hat{g}^{ml}\hat{R}_{ml} =$$
$$-\frac{1}{4}[(-\hat{g})^{1/2}\hat{g}^{kl}(\partial_l\hat{\varphi}^+\cdot\partial_i\hat{\varphi} + \partial_i\hat{\varphi}^+\partial_l\hat{\varphi}) - \delta_i^k(-\hat{g})^{1/2}\hat{g}^{lm}\partial_l\hat{\varphi}^+\cdot\partial_m\hat{\varphi}] \quad (60)$$

$$\partial_i(-\hat{g})^{1/2}\hat{g}^{ik}\partial_k\hat{\varphi} = 0, \quad \partial_i(-\hat{g})^{1/2}\hat{g}^{ik}\partial_k\hat{\varphi}^+ = 0 \quad (61)$$

The work with these new equations is supposed to be conducted with the Hamiltonian gauge

$$(-\hat{g})^{1/2}\hat{g}^{i0} = \delta_0^i \quad (62)$$

Thus, quantum gravity in the Heisenberg representation is a theory of tensor gravitational field $\hat{\psi}_i^k$ and scalar antigravity field $\hat{\varphi}$ (see [7] for details). In what follows, we will continue the anti-gravitating field $\hat{\varphi}$ to call ghosts. From a formal mathematical point of view, Equations (60)–(62)

"do not remember" their origin, therefore the ghosts in them have the status of a physical subsystem dynamically equal rights with a subsystem of gravitons. To return the field $\varphi$ to the standard status of FP ghosts is possible only by the rules of selection of physical states—the postulate that the only admissible states of anti-gravity ghosts are vacuum states with zero occupation numbers. In cases where this is allowed by the properties of the vacuum itself (in the S-matrix theory and in the theory of short-wave gravitons on a slightly curved space-time background), there are no problems with the interpretation of the theory. In a real universe containing a huge (macroscopic) number of gravitons, the situation is different. Let us proceed to extrapolate the theory (60)–(62) to the Universe as a whole.

*2.4. Extrapolation*

From Equations (60)–(62), the equations for gravitons and ghosts in macroscopic space-time with self-consistent geometry can be obtained by identity transformations. However, this procedure itself assumes that there are Heisenberg state vectors in terms of which it is possible to specify the initial states of quantum fields in the macroscopic curved space-time. In the next section, it will be shown that in the self-consistent theory of gravitons and ghosts in an isotropic Universe, such state vectors actually exist as consistent mathematical objects. The derivation of the equations actually reduces to averaging the normal coordinate of the gravitational field over the Heisenberg state vector. As a result of this operation, a $\hat{\psi}_i^{\,k} = \hat{\Psi}_i^{\,k} - \Phi_i^{\,k}$ non-zero C-number function is obtained. A quantum fluctuation is defined as the difference between the operator and its average value $\hat{\psi}_i^{\,k} = \hat{\Psi}_i^{\,k} - \Phi_i^{\,k}$. Substitution of $\hat{\Psi}_i^{\,k} = \Phi_i^{\,k} + \hat{\psi}_i^{\,k}$ into (3) gives

$$(-g)^{1/2} g^{ik} = (-\bar{g})^{1/2} \bar{g}^{il} [\exp(\Psi + \hat{\psi})]_l^{\,k} = (-g)^{1/2} g^{il} (\exp\hat{\psi})_l^{\,k}$$

$$(-g)^{1/2} g^{ik} = (-\bar{g})^{1/2} \bar{g}^{im} (\exp\Phi)_m^{\,k} \tag{63}$$

$$(\exp\hat{\psi})_l^{\,k} = \delta_l^{\,k} + \hat{\psi}_l^{\,k} + \frac{1}{2}\hat{\psi}_l^{\,m}\hat{\psi}_m^{\,k} + ...)$$

We will name the matrix exponent built by C-numerical function $\Phi_i^{\,k}$ the density of contravariant metrics of macroscopic space-time. In such a space, the covariant metric $g_{ik}$, connectivity $\Gamma_{ik}^{\,l}$ and curvature $R_{ik}$ are introduced in the usual way and the covariant derivatives are calculated. Quantum fluctuations $\hat{\psi}_l^{\,k}$ are endowed with properties of the tensor in macroscopic space-time. The Hamiltonian gauge (62) is divided into the gauge of the macroscopic metric: $(-g)^{1/2} g^{i0} = \delta_0^{\,i}$ and the calibration of the quantum field $\hat{\psi}_0^{\,i} = 0$.

The averaging of the equations obtained after substituting (43) into (60) yields Einstein's macroscopic equations:

$$R_i^{\,k} - \frac{1}{2}\delta_i^{\,k} R = \kappa <\Psi|T_i^{\,k}|\Psi>$$
$$T_i^{\,k} = T_{i(grav)}^{\,k} + T_{i(ghost)}^{\,k} \tag{64}$$

where $T_i^{\,k}$ is the total energy-momentum tensor of gravitons and ghosts. Equations for gravitons are obtained by subtraction of averaged Equation (64) from the original Equation (60)

$$\frac{1}{2}(\hat{\psi}_{i;l}^{\,k;l} - \hat{\psi}_{l;i}^{\,k;l} - \hat{\psi}_{i;l}^{\,l;k} + \delta_i^{\,k}\hat{\psi}_{m;l}^{\,l;m}) + \hat{\psi}_i^{\,k} R_i^{\,l} - \frac{1}{2}\delta_i^{\,k}\hat{\psi}_l^{\,m} R_m^{\,l} = \kappa(\hat{T}_i^{\,k} - <\Psi|\hat{T}_i^{\,k}|\Psi>) \tag{65}$$

The semicolon in (65) and later denotes covariant derivatives in macroscopic space-time. The exact expression for the energy-momentum tensor without restrictions on the amplitudes and wavelengths of the quantum fields is given in Section IV of our work [7].

The metric of macroscopic space-time, satisfying Equation (64) and appearing as coefficients in the operator equations for gravitons and ghosts has the status of a self-consistent gravitational field that describes collective interactions in a system consisting of a large (macroscopic) number of particles. It should be emphasized that the appearance of macroscopic space-time in a macroscopic system of particles is a non-perturbative effect, and this effect is accurately taken into account in the equations of quantum gravity (64) and (65). The correlation interactions of the quantum fields outside of the approximation of a self-consistent field are taken into account in the process of solution of these equations by the methods of perturbation theory in terms of the amplitude of the gravitational field. The consistency and mathematical consistency of the system of classical and quantum Equations (64) and (65) in any order of perturbation theory is proved in Section II.F of work [3]. In the approximation in which the quantum field interactions are taken into account only through a classical self-consistent field, the operator equations for gravitons and ghosts are linear.

$$\frac{1}{2}(\hat{\psi}_{i;l}^{\ k;l} - \hat{\psi}_{l;i}^{\ k;l} - \hat{\psi}_{i;l}^{\ l;k} + \delta_i^k \hat{\psi}_{m;l}^{\ l;m}) + \hat{\psi}_l^k R_i^l - \frac{1}{2}\delta_i^k \hat{\psi}_l^m R_m^{\ l} = 0 \qquad (66)$$

$$\hat{\varphi}_{;k}^{\ :k} = 0 \qquad \hat{\varphi}_{;k}^{+ :k} = 0 \qquad (67)$$

The explicit forms of energy-momentum tensors for gravitons and ghosts are given in [3, 7]. They read

$$\kappa \hat{T}^k_{i(grav)} = \frac{1}{4}(\hat{\psi}^l_{m;i}\hat{\psi}^{m;k}_l - \frac{1}{2}\hat{\psi}_{;i}\hat{\psi}^{;k} - 2\hat{\psi}^l_{i;m}\hat{\psi}^{m;k}_l) - \frac{1}{8}\delta_i^k(\hat{\psi}^l_{m;n}\hat{\psi}^{m;n}_l - \frac{1}{2}\hat{\psi}_{;n}\hat{\psi}^{;n} - 2\hat{\psi}^l_{n;m}\hat{\psi}^{m;n}_l) \\ -\frac{1}{2}[\hat{\psi}^l_m(\hat{\psi}^{k;m}_i - \hat{\psi}^{mk}_{;i}) - \hat{\psi}^k_m\hat{\psi}^{l;m}_i + \frac{1}{2}\delta_i^k(\hat{\psi}^n_m\hat{\psi}^{l;m}_n + \hat{\psi}^l_m\hat{\psi}^{m;n}_n)]_{;l} - \frac{1}{2}\hat{\psi}^k_m\hat{\psi}^m_l R^l_i + \frac{1}{4}\delta_i^k \hat{\psi}^n_l \hat{\psi}^m_n R^l_m \qquad (68)$$

$$\kappa \hat{T}^k_{i(ghost)} = -\frac{1}{4}\left(\hat{\varphi}^{+;k}\hat{\varphi}_{;i} + \hat{\varphi}^+_{;i}\hat{\varphi}^{;k} - \delta_i^k \hat{\varphi}^{+;l}\hat{\varphi}_{;l}\right) \qquad (69)$$

The presence of an antigravitating scalar field with the energy-momentum tensor (69) on the right-hand side of the Einstein macroscopic Equation (64) may seem to be a mathematical artifact that has no physical meaning, or even a consequence of some errors made in obtaining the system of Equations (64), (66) and (67). Such impressions, however, are only an emotional reaction to the antigravity of ghosts. The theory under discussion is the result of identical mathematical transformations of the Faddeev path integral [31] to the equations of quantum gravity in the Heisenberg representation, and then to the equations of the theory of gravitons and ghosts in the curved space-time with self-consistent geometry[8]. In fact, it is the presence of ghosts in the system of Equations (64), (66) and (67) that allows us to give a standard interpretation of the results of the theory in those cases when it is possible to draw analogies between the formulation of problems in the theory of the S-matrix and in a self-consistent theory of gravitons and ghosts. We are talking about the problems of a self-consistent theory, in the solution of which only the adiabatic effect of vacuum polarization is taken into account but the super-adiabatic effect of particle production is not taken into account. In the approximation of a stable graviton-ghost vacuum, the system of Equations (64), (66) and (67) is supplemented by standard rules for selecting physical states. The state of gravitons is given by a distribution over non-zero occupation numbers, and the state of ghosts is assumed to be the vacuum. In this case, the ghosts, like in the theory of the S-matrix, compensate for the vacuum polarization effect of gravitons and simultaneously provide a one-loop finiteness of pure gravity without matter fields. Even more important is the fact that the formally ghost-free theory of gravitons in curved space-time does not exist at all (as a model of one-loop quantum gravity), even in the approximation of a stable vacuum, i.e., in the approximation in which pure quantum gravity without matter fields must exist. The reason for the collapse of the ghost-free one-loop theory" is the divergence off the mass shell, the successive renormalizations of which lead to consecutive redefinitions of the graviton field. A complete mathematical proof of this assertion is contained in Section X of work [3].

---

[8] At our request, L.D. Faddeev read the paper [7], in which the above-mentioned identical transformations are described in sufficient detail. Here is an excerpt from L.D. Faddeev report: "My impression is that it is correct, but not very new. Of course, when the action includes ghosts is written, it will allow canonical interpretation. However, the problem of the formal quantization of Einstein's theory of gravity should be considered as solved in all possible formalisms". Our other colleagues (see acknowledgements in [7] shared a similar opinion. In spite of the above, the paper [3] was rejected by the editorial staff of the PRD based on feedback from reviewers who objected to the presence of the ghost sector in the equations of quantum gravity in the Heisenberg representation. Refusing to recognize the results of obvious mathematical transformations two of the four PRD reviewers wrote that in their opinion, the authors of [3] are not familiar with the fundamentals of quantum field theory. The reply to this comment is the work [7], the review by L.D. Faddeev, as well as the text of this paper.

*2.5. The Problem of the Physical Nature of Ghosts*

Spontaneous particle creation in the curved space-time leads to the fact that the vacuum state of quantum fields with zero occupation numbers becomes physically unrealizable. The important fact is that the spontaneous creation of massless particles takes place in the space-time with arbitrary small curvature. For the particles with wavelengths which is much less than the curvature radius it is possible to introduce the local speed of particle creation [40]. In particular, the speed of creation of scalar massless particles reads

$$(nu^i)_{;i} = \frac{1}{288\pi} \cdot R^2 + \frac{1}{960\pi} \cdot C_{iklm} C^{iklm}$$

where $n$ is particle density; $u^i$ is 4-velocity; and $C_{iklm}$ is the Weyl tensor. This equation follows directly from the Klein-Gordon-Fock equation regardless of the nature of the scalar field. Because of this fact, the equation should be considered as a direct mathematical consequence of the ghost Equation (67). The ghost specific is only in the fact that they antigravitate due to "incorrect" sign of EMT (69). The instability of vacuum of shortwave ghosts can be calculated by this equation. In the so-called quasi-resonance spectral region where graviton and ghost wavelengths are comparable with the 4-curvature radius the effect is much stronger. In such a region, there is no a small parameter allowing to consider the vacuum as stable state. The materialization of ghosts of the long wavelengths is a non-perturbative effect of quantum gravity if the theory (which applies to the macroscopic system of gravitons and ghosts in the curved space-time with the self-consistent geometry) and is obtained by extrapolation of the graviton S-matrix to the macroscopic world.

The effect of materialization of anti-gravitating ghosts in the macroscopic world is contained mathematically in the frame of existing theory but obviously it is in a contradiction with the standard concepts based on the "common sense". The contradiction between the mathematics and "common sense" leads to a discussion affecting the fundamental problems of quantum gravity. Let's us come back to the asymptotic states and stable vacuum. In the theory of S-matrix, the asymptotic states and stable vacuum are introduced axiomatically. It is important that these complementary axioms are consistent with the theory's formalism in the frame of perturbative theory. The path integral formalism takes its final form after localization of measure with the use of the ghost field. In such formalism, the axiom about asymptotic states is formulated in terms of the boundary conditions applied to the manifold of metrics over which the integration is performed. The axiom on vacuum stability allows introducing rules for the selection of physical (gauge invariant) states. Formally, the selection rules exclude the non-vacuum states of ghosts with non-zero occupation numbers from the asymptotic physical states. In the theory claiming to be a description of the Universe, the appearance of a curved space-time is a non-perturbative effect. The consequences of that are the lack of asymptotic states and vacuum instability. The disappearance of two key concepts of the perturbative S-matrix at once raises the question of the validity of the extrapolation. The concept of asymptotic states is closely connected with reference frames used in the theory of S-matrix. Here it is assumed that a finite number of reference bodies on which the reference frame is realized are located at spatial infinity, i.e., the means of observation are outside the interaction region. The boundary conditions, for which the path integral is calculated, correspond to the fast damping of inertia fields. Therefore in the theory of the S-matrix inertial frames of reference actually appear. The means of observation located on these frames of reference do not affect the physical process of graviton scattering. It is a radical change in the status of the reference frame that the physical content of the problem of extrapolating the theory of the S-matrix to the Universe as a whole is connected. Such extrapolation is actually carried out for physical conditions in which it is impossible to introduce inertial reference frames at spatial infinity.

One can hope that the effect of the ghost materialization formally following from a rigorous mathematical analysis given in [7] and the present work will be explained in the future theory of quantum gravity, which does not yet exist. As we will see in Section 2.6, the ghost materialization effect takes place for the solutions of one-loop equations describing the accelerated expansion in real time. The cause of this fact is explained in Section 5.2. Meanwhile, as was already mentioned, the problem of ghost materialization does not exist for the one-loop solutions of the instanton type (Section 3) obtained by the Wick rotation with subsequent analytic continuation to real time. In Section 3 of the paper we will focus on these solutions.

*2.6. One-Loop Approximation*

Equations (64), (66)–(69) form a self-consistent set of equations of one-loop quantum gravity in the empty space-time. For the isotropic and homogeneous model of the Universe one can use the FLRW metric. As was shown in our work [3], after eliminating 3-scalar and 3-vector modes Equations (64), (66)–(69) in such a metric read

$$3H^2 = \kappa \rho_g \tag{70}$$

$$2\dot{H} + 3H^2 = -\kappa p_g \tag{71}$$

$$\kappa \rho_g = \frac{1}{8} \sum_{\mathbf{k}\sigma} < \Psi_g | \dot{\hat{\psi}}^+_{\mathbf{k}\sigma} \dot{\hat{\psi}}_{\mathbf{k}\sigma} + \frac{k^2}{a^2} \hat{\psi}^+_{\mathbf{k}\sigma} \hat{\psi}_{\mathbf{k}\sigma} | \Psi_g >$$
$$-\frac{1}{4} \sum_{\mathbf{k}} < \Psi_{gh} | \dot{\hat{\theta}}^+_{\mathbf{k}} \dot{\hat{\theta}}_{\mathbf{k}} + \frac{k^2}{a^2} \hat{\theta}^+_{\mathbf{k}} \hat{\theta}_{\mathbf{k}} | \Psi_{gh} >, \tag{72}$$

$$\kappa p_g = \frac{1}{8} \sum_{\mathbf{k}\sigma} < \Psi_g | \dot{\hat{\psi}}^+_{\mathbf{k}\sigma} \dot{\hat{\psi}}_{\mathbf{k}\sigma} - \frac{k^2}{3a^2} \hat{\psi}^+_{\mathbf{k}\sigma} \hat{\psi}_{\mathbf{k}\sigma} | \Psi_g >$$
$$-\frac{1}{4} \sum_{\mathbf{k}} < \Psi_{gh} | \dot{\hat{\theta}}^+_{\mathbf{k}} \dot{\hat{\theta}}_{\mathbf{k}} - \frac{k^2}{3a^2} r \hat{\theta}^+_{\mathbf{k}} \hat{\theta}_{\mathbf{k}} | \Psi_{gh} > \tag{73}$$

where $\Psi_g$ and $\Psi_{gh}$ are quantum state vectors of gravitons and ghosts, respectively; $H = \dot{a}/a$; $a(t)$ is the scale factor of a flat FLRW model; $\kappa = 8\pi G$; speed of light $c = 1$; dots are time derivatives; $\sigma$ is the polarization index and superscript "+" denotes complex conjugation. Heisenberg's operator equations for Fourier components of the transverse 3-tensor graviton field $\hat{\psi}_{\mathbf{k}\sigma}$ and Grassman ghost field $\hat{\theta}_{\mathbf{k}}$ are:

$$\ddot{\hat{\psi}}_{\mathbf{k}\sigma} + 3H \dot{\hat{\psi}}_{\mathbf{k}\sigma} + \frac{k^2}{a^2} \hat{\psi}_{\mathbf{k}\sigma} = 0 \tag{74}$$

$$\ddot{\hat{\theta}}_{\mathbf{k}} + 3H\dot{\hat{\theta}}_{\mathbf{k}} + \frac{k^2}{a^2}\hat{\theta}_{\mathbf{k}} = 0 \tag{75}$$

Canonical commutation relations for gravitons and anti-commutation relations for ghosts read

$$\frac{a^3}{4}\left[\dot{\hat{\psi}}^+_{\mathbf{k}\sigma}, \hat{\psi}_{\mathbf{k}\sigma}\right]_- = -i\hbar\delta_{\mathbf{k}\mathbf{k}'}\delta_{\sigma\sigma'},$$

$$\frac{a^3}{8}\left[\dot{\hat{\theta}}^+_{\mathbf{k}}, \hat{\theta}_{\mathbf{k}}\right]_+ = -\frac{a^3}{8}\left[\dot{\hat{\theta}}, \hat{\theta}^+_{\mathbf{k}}\right]_+ = -i\hbar\delta_{\mathbf{k}\mathbf{k}'}. \tag{76}$$

One-loop effects of vacuum polarization and particle creation by the background field are contained in Equations (74) and (75) for gravitons and ghosts. These equations are linear in quantum fields but their coefficients depend on the non-stationary background metric. Correspondingly, in the background Equations (70)–(73) we keep the average values of bilinear forms of quantum fields only. In this model, quantum particles interact through a common self-consistent field only. We took also into account the following definitions.

$$<\Psi|\hat{T}^0_0|\Psi> = \varepsilon_g, \qquad <\Psi|\hat{T}^\beta_\alpha|\Psi> = \frac{\delta^\beta_\alpha}{3}<\Psi|\hat{T}^\gamma_\gamma|\Psi> = -\delta^\beta_\alpha p_g \tag{77}$$

Also we have the following rules of averaging of bilinear forms that are the consequence of homogeneity and isotropy of the background.

$$<\Psi_g|\hat{\psi}^+_{\mathbf{k}\sigma}\hat{\psi}_{\mathbf{k}'\sigma'}|\Psi_g> = <\Psi_g|\hat{\psi}^+_{\mathbf{k}\sigma}\hat{\psi}_{\mathbf{k}\sigma}|\Psi_g>\delta_{\mathbf{k}\mathbf{k}'}\delta_{\sigma\sigma'}, \qquad <\Psi_{gh}|\bar{\theta}_{\mathbf{k}}\theta_{\mathbf{k}'}|\Psi_{gh}> = <\Psi_{gh}|\bar{\theta}_{\mathbf{k}}\theta_{\mathbf{k}}|\Psi_{gh}>\delta_{\mathbf{k}\mathbf{k}'}.$$

The mathematically rigorous method of separating the background and the gravitons, which ensures the existence of the EMT, is based on averaging over graviton polarizations: $<\psi^\beta_\alpha> \equiv 0$ if all polarizations are equivalent in the quantum ensemble. Equations (70)–(75) form a self-consistent set of equations of one-loop quantum gravity for gravitons and the FLRW background. For the first time, three solutions to the set of Equations (70)–(75) were obtained in our work [4]. One of these was de Sitter solution, which reads

$$-p = \rho = \frac{3\hbar H^4 N_g}{8\pi^2} \tag{78}$$

where

$$N_g = <n>(\zeta_g \cos\varphi - \zeta_{gh}\cos\chi) \tag{79}$$

$$\varphi = \varphi_{n_{k\sigma}} - \varphi_{n_{k+1,\sigma}} \qquad \chi = \chi_{n_k} - \chi_{n_{k+1}}$$

The first and second terms in RHS of (79) represent the number of gravitons and ghosts, respectively. The RHS of (79) contains parameters of Poisson distributions for occupation numbers of gravitons and ghosts. Explicit forms of these parameters as well as their physical meanings are given in Sections III.D and V.B of work [3]. Averaging of the parameter (79) over the phases yield $N_g = 0$. Therefore, *the solution under discussion does not exist if the superposition of the phases is random.* The coherence of the quantum ensemble, i.e. the correlation of phases in the quantum superposition of the basic vectors, corresponding to the different occupation numbers, points to the fact that the medium is in the graviton–ghost condensate state. The gravitons are dominant in the condensate if $N_g > 0$, and the ghosts are dominant if $N_g < 0$. We call the second case "ghost materialization". Recall that this solution was obtained in real time (with no use of Wick rotation). Speculations on a possible physical nature of ghosts after extrapolation of S-matrix theory to the Universe as a whole are given in Sections 4 and 5 of this paper. Obviously, this is an open question until a quantum gravity theory will be created. Meanwhile, as we will show in Section 3 the self-consistent instanton solutions to Equations (70)–(75) obtained by the Wick rotation are not accompanied by ghost materialization. For the first time, the de Sitter solution for the early Universe based on conformal anomalies was obtained in work [28]. The nature of the de Sitter state produced by gravitational waves is different because in the present finite theory there is no conformal anomalies (Section 3).

*2.7. Conclusion*

It is shown that in the non-stationary Universe, due to the conformal non-invariance and the zero rest mass of gravitons and ghosts, asymptotic states are absent, and the vacuum is unstable, both in the graviton sector and in the ghost sector. The irremovable quantum effect of the spontaneous production of particles in an unstable vacuum makes it impossible to retain for the ghost fields the status of only auxiliary virtual fields, which they had in the theory of the S-matrix. In the non-stationary Universe, ghosts formally acquire the properties of the second physical subsystem, dynamically equal in rights to the subsystem of gravitons. In real time, in the one-loop approximation, the graviton-ghost system forms coherent quantum condensates that produce the de Sitter expansion of the empty space-time, which is accompanied by the ghost materialization. In Section 3, we will show that the de Sitter state formed by instantons is not accompanied by ghost materialization, and it is consistent with observational data on dark energy and inflation. The important argument in favor of the necessity to transition to the Euclidean space of imaginary time is that the same situation occurs in the classical case. To get the de Sitter state due to backreaction of classical gravitational waves it is also necessary to make the transition to the Euclidean space of imaginary time with the subsequent analytic continuation to real time (Section 1).

## 3. Cosmological Acceleration from Virtual Gravitons

In this section, we show that in the one-loop approximation of quantum gravity the virtual gravitons of superhorizon wavelengths generate de Sitter accelerated expansion of the empty isotropic and homogeneous space-time. Again, to get this exact solution one has to make the transition to Euclidean space of imaginary time and then analytically continue it to real time. In both quantum and classical cases, the key element for the appearance of the Sitter expansion is the transition to the Euclidean space of imaginary time with the subsequent analytic continuation to real time. The only difference is that in the quantum case, gravitons form a quantum coherent condensate, which generates a macroscopic quantum effect of cosmological acceleration.

*3.1. Introduction*

In the first part of this work (Section 1), we showed that the classical gravitational waves are able to generate the de Sitter accelerated expansion of the empty space-time. Technically, this is an instanton solution because it was obtained by transition of equations of the theory to imaginary time (Wick rotation) with the subsequent transition to real time by analytic continuation. To study the backreaction effect of gravitons on the expansion of the empty space-time, it was necessary to create the non-contradictory self-consistent quantum theory of gravitons in the Universe which was done in our works [2–4, 7] and in Section 2 of this paper. In our approach, the quantum cosmology is represented as a theory of gravitons in the macroscopic space-time with self-consistent geometry. The quantum state of gravitons is determined by their interaction with the macroscopic field, and the macroscopic field (background geometry), in turn, depends on the quantum state of gravitons. The background metric and the graviton operator appearing in the self-consistent theory are extracted from the unified gravitational field, which initially satisfies exact equations of quantum gravity. In this third part of the paper (Section 3), we show that in the one-loop approximation of quantum gravity the virtual gravitons are able to form the de Sitter state of empty FLRW space-time in real time (without the use of Wick rotation) but in this exact solution the ghosts are materialized. In the frame of the existing quantum gravity theory, the ghost materialization effect has no reasonable interpretation although in the future theory of quantum gravity it will probably have its own interpretation (Section 2). Meanwhile, the de Sitter instanton solution for gravitons obtained by the Wick rotation with the subsequent transition to real time is not accompanied by ghost materialization. For this reason, we consider here the de Sitter instanton solution generated by gravitons. In the end, we show that dark energy effect and inflation generated by gravitational waves and gravitons are consistent with the existing observational data on dark energy and inflation. At the start and by the end of the Universe evolution it is presumed to be empty, so that our analysis is applicable to these extreme states of our space-time. The energy density of the contemporary Universe is decreasing with time as $\rho \sim a(t)^{-3}, t \to \infty$ ($a(t)$ is the scale factor), so that it is going to become empty asymptotically. The contemporary Universe is about 70% empty, so the effects due to emptiness should be noticeable. As we show the dark energy is such an effect caused by gravitational waves and gravitons in the emptying Universe. The still hypothetical inflation can also be caused by gravitational waves and gravitons if the Universe was empty at the start of its evolution which seems very likely. In any case, we show that the observational data on CMB anisotropy and spectrum of fluctuations are consistent with the gravitational wave/graviton theory of inflation.

*3.2. De Sitter State from Gravitons*

In Equations (70)–(75), it is convenient to use the conformal time $\eta = \int dt/a$ and to go from summation to integration by the transformation $\sum_{\mathbf{k}} \ldots \to \int d^3k/(2\pi)^3 \ldots = \int_0^\infty k^2 dk/2\pi^2 \ldots$. From (70)–(75) follow the Friedmanian equations for the energy density and pressure [3]

$$3\frac{a'^2}{a^4} = \kappa \rho_g = \frac{1}{16\pi^2} \int_0^\infty \frac{k^2}{a^2} dk (\sum_\sigma <\Psi_g | \hat{\psi}'^+_{\mathbf{k}\sigma} \hat{\psi}'_{\mathbf{k}\sigma} + k^2 \hat{\psi}^+_{\mathbf{k}\sigma} \hat{\psi}_{\mathbf{k}\sigma} | \Psi_g > $$
$$-2 <\Psi_{gh} | \bar{\theta}'_{\mathbf{k}} \theta'_{\mathbf{k}} + k^2 \bar{\theta}_{\mathbf{k}} \theta_{\mathbf{k}} | \Psi_{gh} >) \tag{80}$$

$$2\frac{a''}{a^3} - \frac{a'^2}{a^4} = -\kappa p_g = -\frac{1}{16\pi^2} \int_0^\infty \frac{k^2}{a^2} dk (\sum_\sigma <\Psi_g | \hat{\psi}'^+_{\mathbf{k}\sigma} \hat{\psi}'_{\mathbf{k}\sigma}$$
$$-\frac{k^2}{3} \hat{\psi}^+_{\mathbf{k}\sigma} \hat{\psi}_{\mathbf{k}\sigma} | \Psi_g > -2 <\Psi_{gh} | \bar{\theta}'_{\mathbf{k}} \theta'_{\mathbf{k}} - \frac{k^2}{3} \bar{\theta}_{\mathbf{k}} \theta_{\mathbf{k}} | \Psi_{gh} >)$$

$$\hat{\phi}''_{k,\sigma} + (k^2 - \frac{a''}{a})\hat{\phi}_{k,\sigma} = 0, \quad \hat{\psi}_{\mathbf{k}\sigma} = \frac{1}{a}\hat{\phi}_{\mathbf{k}\sigma} \tag{81}$$

$$\hat{\vartheta}''_{\mathbf{k}} + (k^2 - \frac{a''}{a})\hat{\vartheta}_{\mathbf{k}} = 0, \quad \hat{\theta}_{\mathbf{k}} = \frac{1}{a}\hat{\vartheta}_{\mathbf{k}}$$

Primes are derivatives over conformal time $\eta = \int dt/a$. The de Sitter background is

$$a = -(H\eta)^{-1} \tag{82}$$

Solutions to Equation (81) over the de Sitter background (82) and the commutation/anticommutation relations for the operator constants (76) read

$$\hat{\psi}_{\mathbf{k}\sigma} = \frac{1}{a}\sqrt{\frac{2\kappa\hbar}{k}}\left[\hat{c}_{\mathbf{k}\sigma} f(x) + \hat{c}^+_{-\mathbf{k}-\sigma} f^+(x)\right],$$
$$\hat{\theta}_{\mathbf{k}} = \frac{1}{a}\sqrt{\frac{2\kappa\hbar}{k}}\left[\hat{\alpha}_{\mathbf{k}} f(x) + \hat{\beta}^+_{-\mathbf{k}} f^+(x)\right], \quad f(x) = (1-\frac{i}{x})e^{-ix} \quad x = k\eta \quad ? \tag{83}$$

$$[\hat{c}_{\mathbf{k}\sigma}, \hat{c}^+_{\mathbf{k}'\sigma}]_- = \delta_{\mathbf{k}\mathbf{k}'}\delta_{\sigma\sigma'}$$

$$[\hat{\alpha}_{\mathbf{k}}, \hat{\bar{\alpha}}_{\mathbf{k}'}]_+ = \delta_{\mathbf{k}\mathbf{k}'} \qquad [\hat{\beta}_{\mathbf{k}}, \hat{\beta}^+_{\mathbf{k}'}]_+ = -\delta_{\mathbf{k}\mathbf{k}'}$$

In accordance with (83), the operator of occupation numbers $\hat{n}_{\mathbf{k}\sigma} = \hat{c}^+_{\mathbf{k}\sigma}\hat{c}_{\mathbf{k}\sigma}$ exists and gives rise to basic vectors $|\hat{n}_{\mathbf{k}\sigma}>$ of Fock's space. Non-negative integer numbers $n_{\mathbf{k}\sigma} = 0, 1, 2, ...$ are eigenvalues of this operator. In our work [4], it was shown that the de Sitter state is an exact self-consistent solution to the set of Equations (80)–(83) in real time. In our work [3], it was shown that this solution can be accompanied by the ghost materialization. The treatment of graviton and ghost state vectors and observables is given in our works [3, 4] (sections III.C and III.D in [3]).

The self-consistent set of equations of the one-loop quantum gravity, which are finite off the mass shell, is formed by (80)–(83). As was shown in Section 2 in general terms, which one also can see from (80)–(83) that in the mathematical formalism of the theory, the ghosts play a role of a second physical subsystem, whose average contributions to the macroscopic Einstein equations are on an equal footing with the average contribution of gravitons. As was already mentioned in Section 2, at first glance, it may seem that the status of the ghosts as the second subsystem is in contradiction with the well-known fact that the Faddeev-Popov ghosts are not physical particles. However the paradox is in the fact that we have no contradiction with the standard concepts of quantum theory of gauge fields but rather full agreement with these. The Faddeev-Popov ghosts are indeed not physical particles in a quantum-field sense, that is, they are not particles that are in the asymptotic states whose energy and momentum are connected by a definite relation. Such ghosts are nowhere to be found on the pages of our work. We discuss only virtual gravitons and virtual ghosts that exist in the area of interaction. As to virtual ghosts, they cannot be eliminated in principle due to the lack of ghost-free gauges in quantum gravity. In the strict mathematical sense, the non-stationary Universe as a whole is a region of interaction, and, formally speaking, there are no real gravitons and ghosts in it although in the short wave approximation $k\eta \gg 1$ gravitons can be considered as real [3] (see more in Section 2).

The set of Equations (80)–(83) can be represented in an alternative form as the Bogoliubov-Born-Green-Kirkwood-Yvon hierarchy or BBGKY chain [3, 4]. To build the BBGKY chain, one needs to introduce the graviton spectral function $W_{\mathbf{k}}$ and its moments $W_n$

$$W_{\mathbf{k}} = \sum_{\sigma} <\Psi_g | \hat{\psi}^+_{\mathbf{k}\sigma}\hat{\psi}_{\mathbf{k}\sigma} | \Psi_g> -2 <\Psi_{gh} | \hat{\theta}^+_{\mathbf{k}}\hat{\theta}_{\mathbf{k}} | \Psi_{gh}>$$

$$W_n = \sum_{\mathbf{k}} \frac{k^{2n}}{a^{2n}} \left( \sum_{\sigma} <\Psi_g | \hat{\psi}^+_{\mathbf{k}\sigma}\hat{\psi}_{\mathbf{k}\sigma} | \Psi_g> -2 <\Psi_{gh} | \hat{\theta}^+_{\mathbf{k}}\hat{\theta}_{\mathbf{k}} | \Psi_{gh}> \right) \qquad n = 0, 1, 2, ..., \infty$$

(84)

The derivation of BBGKY chain can be found in Section V of work [3]. It reads

$$\dot{D} + 6HD + 4\dot{W}_1 + 16HW_1 = 0 \tag{85}$$

$$\dddot{W}_n + 3(2n+3)H\ddot{W}_n + 3\left[\left(4n^2 + 12n + 6\right)H^2 + (2n+1)\dot{H}\right]\dot{W}_n +$$
$$+2n\left[2\left(2n^2 + 9n + 9\right)H^3 + 6(n+2)H\dot{H} + \ddot{H}\right]W_n \qquad n = 1, ..., \infty$$
$$+4\dot{W}_{n+1} + 8(n+2)HW_{n+1} = 0 \tag{86}$$

$$D = \ddot{W}_0 + 3H\dot{W}_0$$

Equations (85) and (86) form the BBGKY chain. Each equation of this chain connects the neighboring moments. Equations (85) and (86) have to be solved jointly with the Einstein Equations (70) and (71). In terms of $D$ and $W_1$ the energy density and pressure of gravitons are

$$3H^2 = \kappa \rho_g \equiv \frac{1}{16}D + \frac{1}{4}W_1, \qquad 2\dot{H} + 3H^2 = -\kappa p_g \equiv \frac{1}{16}D + \frac{1}{12}W_1, \tag{87}$$

Instead of the original self-consistent system of (80) and (81) we get now the new self-consistent set of equations consisting of the Einstein Equation (87) and BBGKY chain (85) and (86). The energy-momentum tensor (87) can be reduced to the form found in [22] by identity transformations. As it was shown in our work [4], the system of Equations (85)–(87) has three exact self-consistent solutions. Two of them are following:

$$D = -48\left[\pm\frac{K^2}{a^2}\left(\ln\frac{a}{a_0}+\frac{1}{2}\right)-\frac{C}{a^6}\right], \qquad W_1 = \pm 24\frac{K^2}{a^2}\left(\ln\frac{a}{a_0}+\frac{1}{4}\right),$$

$$W_m = -24(\mp 1)^m \frac{K^{2m}}{a^{2m}}\ln\frac{a}{a_0}, \qquad m = 2,3,\ldots\infty \tag{88}$$

$$H^2 = \pm\frac{K^2}{a^2}\ln\frac{a}{a_0}+\frac{C}{a^6},$$

where $K^2, a_0, C$ are an arbitrary constants. The analysis of these solutions is given in our work [3]. As was shown in [4], the de Sitter solution is one of exact solutions to the equations of BBGKY chain for the empty FLRW space. One can check by simple substitution that it reads

$$H^2 = \frac{1}{36}W_1 \quad \varepsilon_g = -p_g = \frac{1}{12\kappa}W_1 \quad D = -\frac{8}{3}W_1, \quad a = a_0 e^{Ht} \tag{89}$$

$$W_{n+1} = -\frac{n(2n+3)(n+3)}{2(n+2)}H^2 W_n, \qquad n \geq 1 \tag{90}$$

As is shown in [4], the solution (89) and (90) can be obtained directly from (80) and (81), and de Sitter solution (89) and (90) can be rewritten in the following form

$$D = -\frac{12\hbar N}{\pi^2}H^4, \quad W_n = \frac{(-1)^{n+1}}{2^{2n}}(2n-1)!(2n+1)(n+2)\times\frac{2\hbar N}{\pi^2}H^{2n+2}, \qquad n \geq 1, \tag{91}$$

$$H = const; \quad W_n = const; \quad D = const$$

The zero moment $W_0$, which has an infrared logarithmic singularity, is not contained in the expressions for the macroscopic observables, and for that reason, is not calculated. In the equation for $W_0$, the functions are differentiated in the integrand and the derivatives are combined in accordance with the definition (87). At the last step the integrals that are calculated, already possess no singularities. Both (90) and (91) show that signs of the BBGKY moments alternate because $W_{n+1}/W_n < 0$. This alternation means that the even moments are negative. In accordance with the definition of moments (84), this means that for the even moments we have the following

$$W_n = \sum_{\mathbf{k}}\frac{k^{2n}}{a^{2n}}\left(2<\Psi_{gh}|\hat{\theta}^+_{\mathbf{k}}\hat{\theta}_{\mathbf{k}}|\Psi_{gh}>-\sum_{\sigma}<\Psi_g|\hat{\psi}^+_{\mathbf{k}\sigma}\hat{\psi}_{\mathbf{k}\sigma}|\Psi_g>\right) \tag{92}$$

$$n = 2m; \; m = 1,2,\ldots,\infty$$

Equation (92) shows that the ghosts do not renormalize gravitons, but on the contrary, gravitons renormalize ghosts in even moments. In other words, ghosts begin to play a dominant role, and gravitons begin to play an auxiliary role. This effect we call the ghost materialization. This is the direct confirmation of the fact (discussed in Section 2 and above) that the ghost materialization effect takes place in real time. It is easy to see that a transition to imaginary time (Wick rotation) removes alternation of the moments in (90). This fact allows the expectation of obtaining a de Sitter state from gravitons without ghost materialization.

The imaginary time formalism for the graviton-ghost system was constructed in Section VII of our work [3]. It is fully applicable to the current consideration except for one important problem, which is a method of analytical continuation of solutions obtained for the Euclidean space of imaginary time to the Lorentzian space of real time. The theory is formulated in the space with the metric

$$ds^2 = -d\tau^2 - a^2(\tau)(dx_1^2 + dx_2^2 + dx_3^2) \tag{93}$$

*Operators* of graviton and ghost fields with nontrivial commutation properties are defined over the space (93). Symmetry properties of space (93) allow us to define the Fourier images of the operators by coordinates $x_1, x_2, x_3$, and to formulate the canonical commutation relations in terms of derivatives of operators with respect to the imaginary time $\tau$:

$$\frac{a^3}{4\kappa}\left[\frac{d\hat{\psi}^+_{\mathbf{k}\sigma}}{d\tau}, \hat{\psi}_{\mathbf{k}'\sigma'}\right]_{-} = -i\hbar\delta_{\mathbf{kk}'}\delta_{\sigma\sigma'} \tag{94}$$

$$\frac{a^3}{4\kappa}\left[\frac{d\hat{\vartheta}^+_{\mathbf{k}}}{d\tau}, \hat{\vartheta}_{\mathbf{k}'}\right]_{-} = -i\hbar\delta_{\mathbf{kk}'}, \qquad \frac{a^3}{4\kappa}\left[\frac{d\hat{\vartheta}_{\mathbf{k}}}{d\tau}, \hat{\vartheta}^+_{\mathbf{k}'}\right]_{-} = -i\hbar\delta_{\mathbf{kk}'} \tag{95}$$

Note that (94) and (95) are introduced by the newly independent postulate of the theory, and not derived from standard commutation relations (76) by conversion of $t \to i\tau$. (Such a conversion would lead to the disappearance of the imaginary unit from the right hand sides of the commutation relations.) Thus, the imaginary time formalism cannot be regarded simply as another way to describe the graviton and ghost fields, i.e., as a mathematically equivalent way for real time description. In this formalism the new specific class of quantum phenomena is studied (see footnote 9). The transition to imaginary time $\tau$ and imaginary conformal time $\upsilon$ (Wick rotation) reads

$$t = i\tau \qquad \eta = i\upsilon \tag{96}$$

As follow from (96), the Hubble function in real time $H$ and Hubble function in imaginary time $H_\tau$ are connected by the following relation

$$H = \frac{1}{a}\frac{da}{dt} = -i\frac{1}{a}\frac{da}{d\tau} = -iH_\tau \tag{97}$$

The substitution of (97) into (90) removes the sign alternation. For the first time, the solution of Equations (81) and (82) in real time was obtained in our work [4] where the de Sitter solution (89)–(91) was obtained directly from (80) and (81) For the first time, the de Sitter solution in imaginary time was also obtained from (80) and (81) in our work [2]. The transition to the imaginary time (96) transforms (80)–(82) to the following system

$$-3\frac{a'^2}{a^4} = \frac{1}{16\pi^2}\int_0^\infty \frac{k^2}{a^2}dk(\sum_\sigma <\Psi_g| -\hat{\psi}'^+_{\mathbf{k}\sigma}\hat{\psi}'_{\mathbf{k}\sigma} + k^2\hat{\psi}^+_{\mathbf{k}\sigma}\hat{\psi}_{\mathbf{k}\sigma}|\Psi_g> \tag{98}$$

$$-2<\Psi_{gh}|-\bar{\theta}'_{\mathbf{k}}\theta'_{\mathbf{k}} + k^2\bar{\theta}_{\mathbf{k}}\theta_{\mathbf{k}}|\Psi_{gh}>)$$

$$\hat{\phi}''_{\vec{k},\sigma} - (k^2 + \frac{a''}{a})\hat{\phi}_{\vec{k},\sigma} = 0 \tag{99}$$

$$\hat{\vartheta}''_{\vec{k}} - (k^2 + \frac{a''}{a})\hat{\vartheta}_{\vec{k}} = 0 \tag{100}$$

The de Sitter background in imaginary time reads

$$a = -(H_\tau \upsilon)^{-1} \tag{101}$$

Primes in these equations denote derivatives over imaginary conformal time $\upsilon$. After such transition, solutions to (99) and (100) over the De Sitter background (101) read

$$\hat{\psi}_{\mathbf{k}\sigma} = \frac{1}{a}\sqrt{\frac{2\kappa\hbar}{k}}\left(\hat{Q}_{\mathbf{k}\sigma}g_k + \hat{P}_{\mathbf{k}\sigma}h_k\right), \quad \hat{\theta}_{\mathbf{k}} = \frac{1}{a}\sqrt{\frac{2\kappa\hbar}{k}}\left(\hat{q}_{\mathbf{k}}g_k + \hat{p}_{\mathbf{k}}h_k\right) \tag{102}$$

where

$$g(\xi) = \left(1 + \frac{1}{\xi}\right)e^{-\xi}, \quad h(\xi) = \left(1 - \frac{1}{\xi}\right)e^{\xi}, \quad \xi = k\upsilon \tag{103}$$

The requirement of finiteness eliminates the $h-$ solution, i.e., $\hat{P}_{\mathbf{k}\sigma} = \hat{p}_{\mathbf{k}} = 0$. This requirement leads to the fact that the graviton-ghost system forms a quantum coherent instanton condensate (see Section VII.B of our work [3][9]. The operator functions (102) can be named quantum fields of gravitational instantons of graviton and ghost type or for short, graviton-ghost instantons. Substitution of (102) and (103) into the right-hand-side of (98) leads to

$$3\frac{a'^2}{a^4} = \frac{\kappa\hbar H_\tau^4}{2\pi^2}\int_0^\infty N_k[\xi^2 - (1+\xi)^2]e^{-2\xi}\xi d\xi \tag{104}$$

$$N_k = \sum_\sigma <\Psi_g|\hat{Q}^+_{\mathbf{k}\sigma}\hat{Q}_{\mathbf{k}\sigma}|\Psi_g> - 2<\Psi_{gh}|\hat{q}^+_{\mathbf{k}}\hat{q}_{\mathbf{k}}|\Psi_{gh}> \tag{105}$$

Over the De Sitter background (101), the left-hand-side of (104) must be $3H_\tau^2 = const$ which means that the right-hand-side of (104) cannot be a function of $\upsilon$. In turn, this means that only flat spectrum $N_k = const = N_g$ is able to provide constancy of the right-hand-side of (104).

From (104) and (105) one gets the solution to Equations (98)–(100) in imaginary time. It reads

$$H_\tau^2(1 + \frac{\kappa\hbar H_\tau^2 N_g}{8\pi^2}) = 0 \tag{106}$$

The next step is the analytic continuation of the solution (106) into the Lorentzian space of real time. The analytic continuation of imaginary time solutions (102) and (103) from the Euclidean space into the Lorentzian space of real time can be done by the same way as it was done for the classical gravitational waves (see Equations (21) and (22)). The operators and their derivatives must be continuous at the "barrier" $x = \xi = 0$.

---

[9] As we already mentioned in our work [2], "the effect of dark energy is observed in the contemporary Universe which is far from the Planck time. Therefore, quantum origin of it seems counterintuitive. In fact, this is a macroscopic quantum effect similar to superconductivity and superfluidity [3]. Its origin is related to the formation of quantum coherent condensate. Due to the one-loop finiteness of self-consistent theory of gravitons, all observables are formed by the difference between graviton and ghost contributions. This fact can be seen from the definition of $N_k$. The same final differences of contributions may correspond to the totally different graviton and ghost contributions themselves. All quantum states are degenerate with respect to mutually consistent transformations of gravitons and ghost's occupation numbers, but providing unchanged values of observable quantities. *This is a direct consequence of the internal mathematical structure of the self-consistent theory of gravitons, satisfying the one-loop finiteness condition.* The tunneling that unites degenerate quantum states into a single quantum state produces a quantum coherent instanton condensate in imaginary time which can be analytically continued to real time. We refer the reader to Section VII of our work [3] for details. In a general form, hypotheses on the possibility of graviton condensate formation in the Universe were proposed in [41, 42] (see [3] for more details)".

$$\hat{\psi}_{k\sigma}(\xi=0) = \hat{\psi}_{k\sigma}(x=0) \quad \hat{\theta}_{\mathbf{k}}(\xi=0) = \hat{\theta}_{\mathbf{k}}(x=0)$$
$$\hat{\psi}'_{k\sigma}(\xi=0) = \hat{\psi}'_{k\sigma}(x=0) \quad \hat{\theta}'_{\mathbf{k}}(\xi=0) = \hat{\theta}'_{\mathbf{k}}(x=0)$$
(107)

Same as in the classical case, this procedure allows to express the operator constants, $\hat{Q}_{\mathbf{k}\sigma}$, $\hat{Q}^+_{\mathbf{k}\sigma}$, $\hat{q}_{\mathbf{k}}$ and $\hat{q}^+_{\mathbf{k}}$ in imaginary time through the operator constants $\hat{c}_{\mathbf{k}\sigma}$, $\hat{c}^+_{-\mathbf{k}-\sigma}$, $\hat{\alpha}_{\mathbf{k}}$ and $\hat{\beta}^+_{-\mathbf{k}}$ in real time. As it follows from (107), one gets the following equations

$$\hat{Q}_{\mathbf{k}\sigma} = \frac{\hat{c}_{\mathbf{k}\sigma} - \hat{c}^+_{-\mathbf{k}-\sigma}}{2}$$
$$\hat{q}_{\mathbf{k}} = \frac{\hat{\alpha}_{\mathbf{k}} - \hat{\beta}^+_{-\mathbf{k}}}{2}$$
(108)

The combinations of operator constants from the RHS of (108) form operators of occupation numbers of gravitons, ghosts and anti-ghosts in real time. The imaginary time formalism we presented in Section VII of our work [3]. In combination with the quantum theory of the state vector of the general form (see Sections III.C and III.D of work [3]) it allows to express $N_k$ from (105) through graviton and ghost occupation numbers in real time $<n_{k(g)}>$ and $<n_{k(gh)}>$, which are approximately equal to the number of gravitons and ghosts, respectively (the explicit forms of occupation numbers are presented in [3]). To make the result more transparent, assume that the spectrum is flat and the number of ghosts and anti-ghosts are equal to each other, i.e. $<n_{k(g)}>=<n_g>$; $<n_{k(gh)}>=<\bar{n}_{gh}>=<n_{gh}>$. Assume also that typical occupation numbers in the ensemble are large, so that squares of modules of probability amplitudes are likely to be described by Poisson distributions. In such a case, we get a simple physically transparent result [2,3]

$$N_k = (<n_g> - <n_{gh}>) = N_g = const$$
(109)

where $N_g$ is approximately equal to the true number of gravitons in the Universe[10]. In accordance with (96), we have $H_\tau^2 = -H^2$. Thus, the analytic continuation of (106) to real time reads

$$H^2 = \frac{8\pi^2}{\kappa\hbar(<n_g> - <n_{gh}>)} \text{ if } H^2 \neq 0$$
(110)

$$H^2 = 0, \text{ if } H^2 \neq \frac{8\pi^2}{\kappa\hbar(<n_g> - <n_{gh}>)}$$

As was noted in Section 2 of this paper, in real time ghosts are fictitious particles, which appear to compensate for the spurious effect of vacuum polarization of fictitious fields of inertia. In real time, the gravitational effect of gravitons is expected to be renormalized (decreased) by ghosts exactly as it follows from (110)[11].

---

[10] True in the sense that the contribution of the effect of vacuum polarization of fictitious inertia fields is removed from the gravitational effect of gravitons by renormalization of the number of gravitons

[11] Recall again that despite the widespread belief that the ghost's contribution should not appear in the final result of the calculations, in fact, this is only true for the asymptotic states as it takes place in the S-matrix theory. There are no asymptotic states in the Universe which as a whole is the region of interactions. This is the reason why we deal with virtual gravitons (strictly speaking, there are no real gravitons in the Universe). Ghosts appear here to compensate the effect of vacuum polarization of fictitious fields of inertia [2]. This is the reason why they appear only as a factor renormalizing occupation numbers of gravitons.

As one can see from above, we made the analytic continuation of the imaginary time solution (106) to the real time Lorentzian space by the transition $\tau = -it$, i.e., by the reverse Wick rotation. In our work [3], the analytic continuation was done in a different way. Below is a quotation from Section VII of our work [3]. "Construction of the formalism of the theory is completed by developing a procedure to transfer the results of the study of instantons to real time. It is clear that this procedure is required to match the theory with the experimental data, i.e., to explain the past and predict the future of the Universe. As already noted, the procedure of transition to real time is not an inverse Wick rotation. This is particularly evident in the quantum theory: in (94) and (95) the reverse Wick turn leads to the commutation relations for non-Hermitian operators, which cannot be used to describe the graviton field. The procedure for the transition to real time has the status of an independent theory postulates". The postulate made in [3] leads in particular to the fact that the imaginary time solution (106) is identical to a real time solution, i.e., they are the same. This fact leads immediately to the ghost materialization because in such a case $N_g < 0$ in Equation (106). There is no ghost materialization in (106) if the transition from imaginary to real time is done with reverse Wick rotation as it is done in this work (see (110) and (20), (23) in Section 1 for classical gravitational waves).

The Equation (110) can be rewritten finally in the following form

$$H^2 = \frac{\pi}{G\hbar N_g} \qquad (111)$$

Recall that after the Wick rotation (direct and reverse) the classical gravitational waves also form the de Sitter expansion with the following Hubble function (Equation (24))

$$H^2 = \frac{8\pi}{G <|B|^2>} \qquad (112)$$

where $<|B|^2>^{1/2}$ is the root mean square of action of the ensemble of gravitational waves. Similarly to the classical case, the number of gravitons needed to provide the contemporary Hubble constant (38) is of the order of $N_g \approx 10^{123}$ as it follows from (111). For the first time, this fact was found in [4]. The interpretation of this fact was given in [2,4] (see also Section 4.1.3).

The equation of state of gravitons follows from (104). It is invariant with respect to Wick rotation because of remarkable fact that $H^4 = (-iH_\tau)^4 = H^4{}_\tau$. It reads

$$-p = \rho = \frac{3\hbar H^4 N_g}{8\pi^2} \qquad (113)$$

Thus, the equation of state parameter $w = p/\rho$ in both classical gravitational waves and quantum graviton cases is

$$w = -1 \qquad (114)$$

As is known, quantum gravity cannot be renormalized in higher loops [30]. As also noted above, our derivation was made in the one-loop approximation where it is finite on and off the mass shell. In our work [4], we showed that graviton equation of state (113) can be obtained by a simple qualitative method which does not require discussion of complex nonlinear effects. For the sake of clarity, we give it below. Let us consider the balance of energy that is emerging in space due to graviton creation and disappearance due to graviton annihilation. The characteristic energy of gravitons in these processes is $\hbar\omega \sim \hbar H$. Total probabilities of graviton creation and annihilation (normalized to unity volume) $w_{cr}$ and $w_{ann}$ are proportional to the phase volume of one graviton $\omega^3/3\pi^2 \sim H^3/3\pi^2$. The exponent of the background-graviton coupling constant is unity if $\omega \sim H$. Thus, we obtain for $w_{cr}$ and $w_{ann}$ the following estimates

$$w_{cr} = \frac{\gamma}{3\pi^2} H^3 (\overline{N}_{\mathbf{k}}+1)(\overline{N}_{-\mathbf{k}}+1),$$

$$w_{ann} = \frac{\gamma}{3\pi^2} H^3 \overline{N}_{\mathbf{k}} \overline{N}_{-\mathbf{k}}.$$

Here $\gamma = O(1)$, $\overline{N}_{\pm\mathbf{k}} \sim N_g/2$ is the average number of gravitons with wavelengths that are near the characteristic value $\omega \sim H$. Finally, we get the balance equation in the form.

$$\begin{aligned}\rho_g &= \hbar\omega(w_{cr} - w_{ann}) = \\ &= \frac{\gamma}{3\pi^2}\hbar H^4 (\overline{N}_{\mathbf{k}} + \overline{N}_{-\mathbf{k}} + 1) \simeq \frac{\gamma}{3\pi^2} N_g \hbar H^4\end{aligned} \tag{115}$$

This estimate with accuracy of a numerical factor of the order of unity coincides with Equation (113) which is obtained by exact calculation. Virtual gravitons with wavelength of the order of the horizon must appear and disappear in the graviton vacuum because of massless and conformal non-invariance of the graviton field. A non-zero balance of energy is due to the pure quantum process of spontaneous graviton creation, in other words, due to the uncertainty relation. The permanent creation and annihilation of virtual gravitons is not in exact balance because of the expansion of the Universe. The excess energy comes from the spontaneous process of graviton creation and is trapped by the background.

## 4. Consistency with Observational Data

In this section, we show that the theory of cosmological acceleration from gravitational waves and gravitons is consistent with existing observational data on dark energy and inflation. In case of dark energy, such acceleration provides a transparent explanation of the coincidence problem, threshold problem and the vacuum equation of state of dark energy. Cosmological acceleration by virtual gravitons provides a natural explanation of the origin of inflation, which is consistent with observational data on CMB anisotropy, and spectrum of fluctuations.

*4.1. Dark Energy*

Now, the generally accepted concordance model of cosmological evolution is $\Lambda CDM$ model where $\Lambda$ denotes the cosmological constant and CDM stands for the cold dark matter. As is known the lambda term was introduced by Einstein 100 years ago. Zeldovich [43] proved its connection with the physical vacuum. The equation of state corresponding to lambda term is $p_\Lambda = -\rho_\Lambda$ so that $w_\Lambda = -1$. The numerous observational data (see, e.g., Planck Collaboration data [9]) show that the equation of state parameter of dark energy is $w \approx -1$. Although there are several hypothetical models of dark energy that also predict $w \approx -1$ (see e.g., [44–46] for reviews) the reason for the appearance of $\Lambda$ in the concordance model is the fact that it is not connected with additional hypotheses. Although $\Lambda$-term is consistent with the observational value of $w \approx -1$, there are well-known problems with that. The first is a so-called "old cosmological constant problem": Why is the $\Lambda$-value measured from observations is of the order of $10^{-123}$ vacuum energy density? The second one is a "coincidence problem": Why is the acceleration happening during the contemporary epoch, not earlier and not later? If the attempts to answer the first question can be connected with some speculations about "fine tuning", there is no reasonable ideas to answer the second question (perhaps with the exception of the anthropic principle). In distinction to speculative hypotheses on causes of cosmological acceleration of the contemporary Universe (quintessence and others), the gravitational waves are not hypothetical, they are real and observable [8]. They provide the equation of state parameter $w = -1$. Moreover, only the classical and quantum gravitational waves provide physically transparent answer to the question: Why is the acceleration happening during the contemporary epoch, not earlier and not later? The arguments in favor of classical and quantum gravitational waves as the cause of the dark energy effect presented in this paper allow us to propose a GWCDM model as a concordance model where GW stands for Gravitational Waves.

4.1.1. Coincidence Problem

Up to replacement of operators by C-functions, the equations for gravitational waves and gravitons over the background metric $a(\eta)$ are identical. Those are Equations (9) and (81). We give this equation here again

$$\phi''_{\vec{k},\sigma} + (k^2 - \frac{a''}{a})\phi_{\vec{k},\sigma} = 0 \tag{116}$$

Equation (116) is a Schrödinger-like equation, in which $a''/a$ plays the role of "one-dimensional potential". The only difference is that the spatial coordinate in the Schrödinger equation is changed to the time coordinate in the Schrödinger-like Equation (116). The equation of state of non-relativistic matter filling the contemporary Universe is $p = 0$, the corresponding background scale factor is $a(\eta) = \eta^2$, so its "one-dimensional potential" is $a''/a = 2/\eta^2$. The equation of state of the de Sitter state is $p = -\rho$, the background scale factor for the de Sitter state is $a(\eta) = -(H\eta)^{-1}$, so that the de Sitter "one-dimensional potential" is $a''/a = 2/\eta^2$, which exactly coincides with the "one-dimensional potential" of non-relativistic matter. Thus, gravitational waves and gravitons do not "feel" a difference between the non-relativistic matter with the equation of state $p = 0$ and de Sitter State with the equation of state $p = -\rho$. This coincidence takes place in this and only in this case, i.e., for the non-relativistic matter filling the contemporary Universe and the de Sitter state which is produced by classical and quantum gravitational waves. The coincidence of "one-dimensional potentials" provides the smooth transition from one regime to another only for equations of state $p = 0$ and $p = -\rho$, i.e., only in the contemporary epoch of the Universe evolution. For the first time, this solution to the coincidence problem was proposed in our work [2]. In the frame of such reasoning, one can answer the question why dark energy was unable to appear during the radiation dominated epoch. As is known, the energy equation reads (in this case)

$$3H^2 = 8\pi G(\varepsilon_g + \frac{c_r}{a^4} + \frac{c_m}{a^3}) \tag{117}$$

The de Sitter solution could appear if the last stage of evolution is followed by a vacuum stage. However, the second term of RHS in (117) vanished first, entailing no vacuum but CDM stage of evolution[12]. Only after the disappearance of the non-relativistic matter (the third term of RHS) the vacuum stage can occur and with it the de Sitter state.

4.1.2. The Threshold Problem

There is another unanswered question adjoining to the coincidence problem. Why did the dark energy manifest itself only when the energy densities of non-relativistic matter and dark energy became comparable, i.e., relatively recently? To answer this question, we have to take into account pressureless matter that fills the contemporary Universe. This means that we have to add a term $\kappa\rho_m$ to the RHS of Equation (80) where $\rho_m$ is the energy density of this matter, i.e. $\rho_m = \rho_{m,0}(a_0/a)^3$,. After that, Equation (80) takes the following form

$$3\frac{a'^2}{a^4} = \frac{1}{16\pi^2}\int_0^\infty \frac{k^2}{a^2}dk(\sum_\sigma <\Psi_g|\hat{\psi}'^+_{\mathbf{k}\sigma}\hat{\psi}'_{\mathbf{k}\sigma} + k^2\hat{\psi}^+_{\mathbf{k}\sigma}\hat{\psi}_{\mathbf{k}\sigma}|\Psi_g>$$
$$-2<\Psi_{gh}|\bar{\theta}'_{\mathbf{k}}\theta'_{\mathbf{k}} + k^2\bar{\theta}_{\mathbf{k}}\theta_{\mathbf{k}}|\Psi_{gh}>) + \kappa\rho_0(\frac{a_0}{a})^3 \quad (118)$$

---

[12] In modern cosmology the contemporary epoch of the Universe evolution is commonly called "matter dominated", although in reality it is better to call it "gravitational wave dominated epoch".

Recall that a necessary condition for the de Sitter solution to be obtained from gravitational waves and gravitons is the transition to the imaginary time by Wick rotation with the subsequent analytic continuation to the Lorentz space of real time. We applied this procedure to Equations (1)–(4), which were written for the empty space-time. If the space-time is not empty and contains some matter, then Equation (1) (recall that it is in imaginary time) reads

$$-3\frac{a'^2}{a^4} = \kappa(\rho_{inst} + \rho_m)$$
$$\rho_{inst} = \frac{1}{16\pi^2\kappa}\int_0^\infty \frac{k^2}{a^2}dk(\sum_\sigma <\Psi_g|-\hat{\psi}'^+_{\mathbf{k}\sigma}\hat{\psi}'_{\mathbf{k}\sigma} + k^2\hat{\psi}^+_{\mathbf{k}\sigma}\hat{\psi}_{\mathbf{k}\sigma}|\Psi_g> \quad (119)$$
$$-2<\Psi_{gh}|-\bar{\theta}'_{\mathbf{k}}\theta'_{\mathbf{k}} + k^2\bar{\theta}_{\mathbf{k}}\theta_{\mathbf{k}}|\Psi_{gh}>)$$

where $\rho_{inst}$ is the energy density of instantons, i.e., energy density of quantum metric fluctuations in imaginary time; and $\rho_m$ is the energy density of matter and primes are derivatives over imaginary conformal time (96). Taking into account the fact that $\rho_{inst} < 0$ (see, e.g., (104)), one can rewrite (118) in the following form

$$3\frac{a'^2}{a^4} = \kappa(|\rho_{inst}| - \rho_m) \quad (120)$$

In the case of non-relativistic matter, $\rho_m = \rho_{m,0}(a_0/a)^3$ so that $a \to \infty$, $\rho_m \to 0$. Equation (120) tells us that solutions to Equation (120) can exist only if $\rho_m \leq |\rho_{inst}|$. In other words, the energy density of non-relativistic matter must fall below a certain limit (energy density of instantons) to make it possible for the appearance of the dark energy effect. For the first time, this threshold effect was discovered in our work [2] where its consistency with the observational data was discussed. The threshold effect for both inflation and dark energy was discussed in [1]. Below is a quotation from [1]. "In the dark energy case, the energy density of non-relativistic matter is initially higher than this threshold, i.e., $\rho_m > |\rho_{inst}|$. Therefore, when $\rho_m$ drops to the threshold $\rho_m \leq |\rho_{inst}|$ (and below) it marks the birth of dark energy which is a possible explanation to the fact that the birth of dark energy is taking place during the contemporary epoch of cosmological evolution ("coincidence problem"). In the case of inflation, the situation is reversed in the following sense. Presumably, there was no matter initially, so the Universe started with the natural de Sitter expansion. With time, the newborn matter began to appear; so that de Sitter expansion was changed to quasi-de Sitter, and finally, the inflation stops when the energy density of a newborn matter $\rho_M$ has increased up to the threshold $\rho_M = |\rho_{inst}|$. After that, the standard Big Bang cosmology begins. It is convenient to characterize the difference between the cases of inflation and dark energy by the parameter $\delta = -\dot{H}/H^2$ (so-called "slow-roll" parameter in the scalar field theory) where the sign of $\delta$ indicates the direction of change in $H$. If the Hubble function $H$ is increasing with time then $\delta < 0$ and vice versa. Note that $\delta = \delta_\tau$ where $\delta_\tau = -\dot{H}_\tau/H_\tau^2$ (here the dot is derivative over $\tau$). In the case of DE, one has to expect an increase in $H_\tau$ with time because $\rho_m \sim a^{-3}$ is decreasing, and hence $\delta_{DE} < 0$. In the case of inflation, the energy density of a newborn matter $\rho_M$ is increasing with time (until it stops after it reaches the threshold (3)), so one has to expect $\delta_{\inf} > 0$ in this case. Such behavior of parameter $\delta$ is consistent with observations (see Section 4.2.2)".

Thus, both gravitational waves and gravitons generate the de Sitter expansion of the empty (and almost empty) Universe. The equation of state parameter of dark energy that they produce is $w = -1$ which is consistent with observational data. This mechanism provides a physically transparent explanation of the coincidence and threshold problems.

### 4.1.3. $\Lambda_{vacuum} / \Lambda_{observable} \approx 10^{123}$ Controversy

Cosmological acceleration from gravitons provides a transparent explanation to a "mystic" discrepancy in $10^{123}$ between observed value of $\Lambda$ and its theoretical prediction. This number has nothing to do with $\Lambda$. It is the number of gravitons in the contemporary Universe $N_g \approx 10^{123}$.

### 4.1.4. The Need to Compare Theory with Other Observational Data

The gravitational wave theory allows us to resolve the three main paradoxes of dark energy described above without resorting to the anthropic principle. Nevertheless, it is also necessary to verify that the theory is consistent with other observational data on the dark energy presently available. The calculation was made by us in 2008 ([47], Section IX). We started with a qualitative analytic model consisting of combination of the BBGKY solution (88) and non-relativistic matter $\rho_m$. We evaluated our $\Lambda$GCDM model (where G stands for gravitons) in the comparison with $\Lambda$CDM model. The fitting curves for distant modules of Supernovae SNIa for both models proved to be statistically almost undistinguishable from $\Lambda$CDM in accordance with $\chi^2$ criterion, with a small advantage of $\Lambda$GCDM model over the $\Lambda$CDM model In addition to the Hubble diagram for supernovae SNIa, the information about Dark Energy is contained in the Hubble diagram for radio-galaxies and gamma-ray bursts. It is also contained in the cosmological parameters extracted from the CMB data and correlation functions characterizing the large scale structure of the Universe. We did not use radio-galaxies and gamma-ray bursts data because of big statistical errors of the relevant data. The density of Dark Energy on the cosmological scale from the instant of the last scattering $z_{ls}$ up to the present time is contained in the shift parameter. Acoustic oscillations in the photon–baryon plasma prior to recombination give rise to a peak in the correlation function of galaxies. This effect has been measured in a sample of luminous red galaxies and leads to the value $A = 0.469(n/0.98)^{-0.35} \pm 0.017$ where $n = 0.96$ is the spectral index of the primordial power spectrum. Our model predicted $0.413 < A < 0.479$. When the two models were compared, two facts emerge. First, the application of the $\chi^2$ criterion to the smoothed data of the $\Lambda$GCDM model shows that the model gains a more significant advantage. Second, comparison of Sections IX.17 and IX.20 from [47] shows that on the smoothed data the intervals of statistical sums do not overlap with the lesser value of the statistical sum of the $\Lambda$GCDM model. Therefore, by statistical criteria obtained from the smoothed Hubble diagram for supernovae SNIa, the $\Lambda$GCDM model has an advantage. The appropriate references are given in our work [47] (original version) which exists only as a preprint. It was not published because we were going to consider the exact model instead of this qualitative (toy) model. After adding the energy density of matter to the RHS of Equation (80) we get the system of Equations (81) and (118) which is unsolvable in the analytical form. We have to look for a numerical solution of (81) and (118) in the Euclidean space of imaginary time with subsequent analytic continuation into the Lorentzian space of real time. After the solutions are obtained, we have to compare them with observational data described above. Actually, we have asymptotic solutions for the set of Equation (81) and (118) for $\rho_m \gg \rho_g$ and $\rho_m \ll \rho_g$ [4] but this is not enough for the comparison with observational data described above. Unfortunately, we did not go any further than this qualitative model[13].

*4.2. Inflation*

In distinction to the dark energy effect, which is the established observational fact [13, 14], the idea of the necessity of inflation does not yet have direct reliable observational confirmation. It seems attractive due to the ability to solve three known problems, which are flatness, horizon and monopoles [10]. It is worth recalling Weinberg's remark [48] that "so far, the details of inflation are unknown and the whole idea of inflation remains a speculation, though one that is increasingly plausible". In the case of inflation, it is almost generally accepted that the acceleration of expansion is due to a hypothetical scalar field. Gravitational waves and gravitons are not hypothetical, and they are able to provide the de Sitter acceleration of the empty Universe as was shown in this work. We assume that the Universe was empty before the first matter was born. If so, the theory of cosmological acceleration developed in this work is applicable to inflation also. In this sub-section, we show that inflation from gravitons is consistent with the observational data on CMB anisotropy and spectrum of fluctuations.

4.2.1. CMB Anisotropy from Fluctuations of Number of Gravitons

The following calculation of CMB anisotropy due to fluctuations of number of gravitons in the Universe was presented in our work [1]. In view of the importance of this result, we reproduce it here again. The equation of state (113) was obtained under the assumption that the graviton spectrum is flat and typical occupation numbers in the ensemble are large, so that squares of modules of probability amplitudes are likely to be described by Poisson distribution [3]. If so, from Equation (111) it follows that in the energy units $\hbar = c = 1$, $H$ reads

$$H^2 = 8\pi^2 \frac{M_{pl}^2}{\alpha N_g} \qquad (121)$$

where $M_{pl}$ is the reduced Planck mass $M_{pl} = (\hbar c / 8\pi G)^{1/2} = 2.4 \cdot 10^{18} GeV/c^2$. From (111) it follows that the fluctuations of the number of gravitons $N_g$ that are presumably described by Gaussian distribution are $<(\Delta N_g)^2>/<N_g>^2 = <N_g>^{-1}$, so that

$$\frac{<(\Delta N_g)^2>}{<N_g>^2} = \frac{\alpha}{8\pi^2} \cdot \frac{H^2}{M_{pl}^2} \qquad (122)$$

Here $<(\Delta N_g)^2> = <N_g^2> - <N_g>^2$. Due to Equation (113), one gets

$$\frac{<(\Delta N_g)^2>}{<N_g>^2} = \frac{<(\Delta \rho)^2>}{<\rho>^2} \qquad (123)$$

Thus, fluctuations of the number of gravitons produce fluctuations of energy density. They play the same role as scalar perturbations (density fluctuations) which are responsible for the anisotropy of CMB in the models of inflation using scalar fields. In other words, if gravitons are responsible for the inflation then fluctuations of the number of gravitons are the cause of the anisotropy of CMB. For the typical energy scale of inflation $H \simeq 10^{15} GeV$ and $\alpha = 1$ one gets from (121) and (122).

$$(\frac{<(\Delta \rho)^2>}{<\rho>^2})^{1/2} \simeq 1.5 \cdot 10^{-5} \qquad (124)$$

---

[13] After the death of one of the co-authors (Grigory Vereshkov), this work was suspended.

As is known, temperature fluctuations $\Delta T/T$ are of the same order of magnitude as the metric and density perturbations, which contribute directly to $\Delta T/T$ via the Sachs-Wolfe effect. Thus, the fluctuations of the number of gravitons in the Universe are able to produce CMB anisotropy $\Delta T/T \sim 10^{-5}$ due to fluctuations of gravitational potential which in turn are of the order of fluctuations of energy density.

4.2.2. Spectrum of Metric Fluctuations

As was shown above, the de Sitter solution is produced by the flat spectrum of metric fluctuations. The observed tilt $n_s - 1$ of the power spectrum $k^{n_s - 1}$ deviates slightly from the scale-invariant form corresponding to $n_s = 1$. The observed value is $n_s = 0.96 \pm 0.013$ [49]. This means that in reality we deal with a quasi-de Sitter expansion. As was shown in our work [1], assuming that $N_k = N_0 (k/k_0)^\beta$ ($k_0$ is a pivot scale) we get for $\beta$

$$\beta = 3(1+w) \tag{125}$$

In the case of inflation, one starts from the empty space that is gradually filled with the newborn matter. In case of dark energy, one gets the opposite process, one starts from the space filled with matter that is gradually emptied (see [1] and Section 4.1.2). This leads to the fact that we have to expect to find $w < -1$ for dark energy and $w > -1$ for inflation [1]. In the case of dark energy, the Planck data [9] obtained by a combination of Planck+WP+BAO give, e.g., $w = -1.13^{+0.24}_{-0.25} < -1$. In case of inflation, $n_s - 1 = -\beta$, and for $n_s \approx 0.96$, we get $\beta_{\inf} \approx 0.04$ which leads to $w = -0.987 > -1$ [1]. Thus, the graviton theory of dark energy and inflation predicts the correct sign of the equation of state parameter $w$, which is confirmed by observational data.

**5. Origin of Acceleration**

At the start and by the end of the cosmological evolution the Universe is presumed to be empty (with no matter fields). In the extreme conditions of empty space-time at the start and at the end of the Universe evolution, the energy needed to generate an accelerated expansion comes from the Euclidean space, i.e., from "nothing".

*5.1. Wick Rotation*

As is well known, to solve the non-stationary Schrödinger equation one needs to make the Wick rotation $t = i\tau$, which transforms the Schrödinger equation into a heat equation solutions to which are well known. This fact gave rise to a widespread belief that the Wick rotation is simply a convenient change of variables. As was mentioned in our work [6], the situation with this rotation is not so simple. Recall that the self-consistent sets of equations for classical gravitational waves Equations (7)–(10) have no solutions in real time because of divergences of integrals. The self-consistent solutions to these equations can be obtained in imaginary time only by the Wick rotation and should be analytically continued to real time. Quantum metric fluctuations in imaginary time (graviton-ghost instantons) form a macroscopic de Sitter state in real time as do gravitational waves. As is known, usually the Wick rotation is used in the quantum instanton theories [50]. However, as was shown in Sections 1.2 and 1.3, the Wick rotation leads to the same de Sitter expansion of empty space in classical cases too, and this means that a central role is played not by the quantum nature of the effect but the transition to imaginary time by the Wick rotation.

The best known attempt to ascribe physical meaning to the imaginary time was made by Hartle and Hawking [51]. As is shown in this paper, the de Sitter accelerated expansion of the empty FLRW space under backreaction of classical gravitational waves and/or quantum gravitons require a *mandatory* transition to the Euclidean space of imaginary time and then return to the Lorentzian space of real time. On the other hand, the de Sitter accelerated expansion of the empty space by the end of the evolution of the Universe is confirmed by observational data (dark energy). The hypothesis on inflation seems "increasingly plausible" [48]. One can assume that the very existence of these two effects is evidence to the fact that in the very extreme conditions of empty spacetime (at the start and by the end of the Universe evolution) the Euclidean space is the source of energy for acceleration (see Sections 5.2 and 6).

*5.2. Where Does the Energy Come from?*

In the FLRW metric, the deceleration/acceleration expansion is described by the Friedman equation

$$\frac{\ddot{a}}{a} = -\frac{\kappa}{6}(\rho + 3p) \tag{126}$$

The accelerated expansion ($\ddot{a} > 0$) can take place only if $p < -\rho/3$. In accordance with thermodynamics, the energy conservation law is $dE = -pdV$. The negative pressure leads to the fact that the total energy $E$ of the Universe is increasing with the increasing of space volume $V$ in the process of expansion. In de Sitter case, $p = -\rho = const$ so that $E = \rho V \sim a(t)^3$, i.e., in the process of expansion the energy of the Universe is increased indefinitely. To keep the energy density constant during the de Sitter accelerated expansion, one needs a source of energy which is external to the space of the Universe, i.e., one needs an "outsourcing". The question is where can such energy come from? As is known, the vacuum Einstein equations with the cosmological constant $\Lambda$ also produce the de Sitter solution in the FLRW metric corresponding to the equation of state $p_\Lambda = -\rho_\Lambda$. A comprehensive review of the evolution of ideas that led to the understanding of $\Lambda$ as a vacuum energy density can be found in the work [52]. The important fact is that the physical vacuum with such an equation of state is an external source of energy for the empty FLRW space-time, and this is the reason why $\Lambda$ is able to produce the accelerated expansion as well as de Sitter gravitational instanton [2] (see below). Other popular sources of external energy are hypothetical scalar fields which are also of vacuum origin.

To answer the questions in the title of this section, one needs to substitute $\rho_g$ and $p_g$ from Equations (72), (73) to Equation (126). This substitution reads

$$\frac{\ddot{a}}{a} = -\frac{1}{12}(\sum_{k\sigma}<\Psi_g|\dot{\psi}^+_{k\sigma}\dot{\psi}_{k\sigma}|\Psi_g> - 2\sum_{k}<\Psi_{gh}|\dot{\theta}^+_k\dot{\theta}_k|\Psi_{gh}>) \tag{127}$$

Recall that in real time, ghosts are fictitious particles, which appear to compensate for the spurious effect of vacuum polarization of fictitious fields of inertia. Therefore, figuratively speaking the observables are formed by differences ("gravitons"-"ghosts"), i.e., exactly like it is seen in Equation (127). In other words, ghosts renormalize (decrease) the gravitational effect of gravitons compensating the spurious effect of vacuum polarization of fictitious fields of inertia. In real time, to get the acceleration $\ddot{a} > 0$, the RHS of (127) must change its sign, i.e., ghosts must prevail over gravitons. Figuratively speaking, we call it "ghost materialization". Therefore, the "ghost materialization" looks like our cost for getting the acceleration. This general conclusion is, of course, confirmed by the explicit de Sitter (accelerated) solution obtained in real time (90)–(92) (see also (78) and (79)). Thus, the ghost subsystem is a source of energy needed to accelerate the expansion in real time. In other words, the energy needed for the acceleration comes again from the vacuum. As was seen in Section 2, the interpretation of this fact is an open question. As a matter of fact, a deeper analysis (in imaginary time) shows that there is no need in "ghost materialization" to get an accelerating expansion from gravitons as it is shown in Section 3.2[14].

The transition to imaginary time (96) means the transition from Lorentzian space of real time to Euclidean space of imaginary time (93). In the Euclidean space of imaginary time, we obtained the de Sitter gravitational instanton (106) as a solution to Equations (98)–(100) (see also our work [2]). As is mentioned in Section 6, in Euclidean space of imaginary time the gravitational waves are damped spending their energy on the formation of de Sitter gravitational instanton. The analytic continuation of this solution to Lorentzian space of real time gives us the accelerated de Sitter expansion in real time (110). In other words, the energy needed to accelerate the expansion in real time comes from Euclidean space, i.e., from "nothing". The same origin of the de Sitter acceleration from "nothing" is valid for the classical gravitational waves (Section 1). In distinction to the quantum graviton case, this is the only a possibility to get the de Sitter state in real time from classical gravitational waves. Note the remarkable fact that in both classical and quantum instanton solutions the equation of state in real time turns out to be vacuum $p = -\rho$ as well as in the case of ghost materialization. Recall that Einstein introduced the cosmological constant in 1917 to make a static Universe. It took 50 years to find out that $\rho_\Lambda$ is the energy density of vacuum [43]. The Wick rotation produces the vacuum equation of state $p = -\rho$ in the Euclidean space of imaginary time, and it is invariant with respect to the transition to Lorentzian space of real time (113). What is the relation of Wick rotation to physical vacuum?

*5.3. Gravitational Waves vs. Scalar Field*

Starting from pioneering works [10–12] and others, the scalar field was considered as a possible cause of inflation. Choosing a different form of the potential of this field, the authors attempt to reconcile the theory with a number of e-foldings needed to ensure the flatness of the modern Universe and to solve the horizon and monopole problems (references to the original works can be found in Weinberg's book [48]) and get an agreement with the CMB observations. In our work [6], we used a diametrically opposite approach to the idea of the scalar field as a possible cause of inflation. We showed that the ensemble of randomly distributed non-interacting scalar fields, which is homogeneous and isotropic on the average with potentials $V = const$ and on the average, homogeneous and isotropic ensembles of classical and quantum gravitational waves generate the same de Sitter expansion of the empty space-time. This fact takes place because the set of equations describing on the average, a homogeneous and isotropic scalar fields with constant potential is identical to on the average, homogeneous and isotropic ensemble of non-interacting classical gravitational waves with lambda term $\Lambda \neq 0$. Such an ensemble of homogeneous and isotropic on the average scalar fields with zero potential is equivalent to such an ensemble of gravitational waves with the zero lambda term ($\Lambda = 0$).

As is known (see, e.g., [53]), the energy density $\rho$ and pressure $p$ of a single real scalar field $\varphi(\vec{x}, t)$ in the FLRW metric read

$$\begin{aligned}\rho &= \frac{\dot\varphi^2}{2} + \frac{(\nabla\varphi)^2}{2a^2} + V(\varphi) \\ p &= \frac{\dot\varphi^2}{2} - \frac{(\nabla\varphi)^2}{6a^2} - V(\varphi)\end{aligned} \quad (128)$$

where $V(\varphi)$ is a potential of the scalar field. The equation of motion for this field reads

---

[14] Recall that the Faddeev-Popov ghosts are fictitious particles, and a possibility of their "materialization" in real time affects the subtle questions of the theory of gauged fields in the non-stationary Universe (Section 2). Because of the present lack of a consistent quantum theory of gravity, it is an open question today. The situation is different in imaginary time which is a "main player" in this work.

$$\ddot{\varphi} + 3H\dot{\varphi} - \frac{1}{a^2}\nabla^2\varphi - \frac{\partial V}{\partial \varphi} = 0 \qquad (129)$$

After averaging over the ensemble of such scalar fields (randomly distributed but homogeneous and isotropic in the average), we get Equations (1)–(5) of Section 1 describing the ensemble of classical gravitational waves (randomly distributed but also homogeneous and isotropic in the average). Although the equations for the scalar field and gravitational waves are formally identical as well as are their solutions, they have a different origin. In the above study, we used equations in which the nonlinear interaction of waves is not taken into account. In other words, equations for gravitational waves are approximate equations; meanwhile equations for a scalar field are exact equations. Therefore, the solutions (24) and (25) of Section 1 are *exact solutions* in the case of randomly distributed (homogeneous and isotropic in the average) scalar fields.

From the other hand, gravitational waves are very experimentally confirmed phenomenon [8] meanwhile a scalar field is still a hypothetical thing. Recall that regardless of the origin of these equations, in order to obtain the de Sitter solution in real time it is necessary to use a Wick rotation.

*5.4. Virtual Gravitons vs. Classical Gravitational Waves*

As one can see from Sections 4.2.1–4.2.2 of this paper, both observed effects, which are associated with inflation, are consistent with the theory of cosmological acceleration caused by gravitons. The inflation (if it exists) should begin soon after the Planck time, so it is natural to assume that the gravitational waves responsible for these effects should be quantum, i.e., gravitons. The situation is not so unambiguous in the case of dark energy. All three observed effects of dark energy described in Sections 4.1.1–4.1.3 are compatible with both approaches, quantum and classical. It is because of the fact that Equations (9) and (81) are valid for both quantum and classical cases, the same reasoning is applicable to both cases. The dark energy effect is observed in the current epoch of cosmological evolution that is far enough from the Planck time, so at first glance the classical gravitational waves seem preferable. Despite the fact that the dark energy effect from quantum gravitons seems counterintuitive in fact, it is possible. The gravitons of super-horizon wavelengths form a quantum coherent instanton condensate which is a macroscopic effect of quantum gravity similar to superconductivity and superfluidity ([3, 4] and Footnote 9). Such a macroscopic effect of quantum gravity may take place in the current epoch of cosmological evolution as well as the acceleration from classical gravitational waves. In any case, as we see the dark energy effect can be caused by gravitational waves, classical and/or quantum.

*5.5. Conclusion*

It is shown that quantum and classical metric fluctuations of the empty space-time (gravitons and classical gravitational waves) lead to the formation of the de Sitter state at the start and by the end of the Universe evolution. Such an accelerated de Sitter expansion of the Universe by the end of its evolution due to gravitational waves is consistent with observational data on dark energy. It explains a coincidence paradox (Why is the acceleration happening during the contemporary epoch, not earlier and not later?), threshold paradox (Why did the dark energy manifest itself only when the energy densities of non-relativistic matter and dark energy became comparable?) and $10^{123}$ controversy (Why $\Lambda_{vacuum}/\Lambda_{observable} \approx 10^{123}$ ?). Such an accelerated expansion of the Universe at the start of its evolution due to virtual gravitons is consistent with the observational data on CMB anisotropy and spectrum fluctuations caused by inflation.

**6. Cosmological Scenario**

In the above study, we started with Lorentzian space of real time, made Wick rotation to Euclidean space of imaginary time where we found the self-consistent de Sitter solution, which was analytically continued back to Lorentzian space of real time. We would have obtained the same result if we started from the Einstein equations written directly for Euclidean space (93), without resorting to the Wick rotation. With this approach, we would again find that the same de Sitter state is the self-consistent solution to the backreaction problem. It is because both approaches lead to the same set of Equations (98)–(100) for gravitons and (13) and (14) for the classical gravitational waves in Euclidean space. Therefore, solutions to these are identical in both approaches. Should we not assume that Euclidean spacetime is the source of energy entering the Lorentz spacetime of the Universe? In such a case, from the point of view of the observer located in the Lorentzian space of the Universe, the energy (needed for the accelerated de Sitter expansion) comes from "nothing" (see, e.g., [2] and references therein).

The mechanism for the appearance of an accelerated expansion of space is as follows. In Lorentzian space of real time both classical and quantum gravitational waves are not damped as one can see from Equations (12) and (83). In Euclidean space of imaginary time both classical and quantum gravitational waves are damped (see (17) and (103)) spending their energy on the formation of de Sitter gravitational instanton as is seen from (17), (18), (103) and (104). We write once again Equations (81) and (99) describing gravitational waves in the Lorentzian space of real time and Euclidean space of imaginary time, respectively. For superhorizon wavelengths $k\eta \ll 1$, which is the case in this work, they read

$$\frac{d^2 \hat{\phi}_{\vec{k},\sigma}}{d\eta^2} - \frac{1}{a}\frac{d^2 a}{d\eta^2} \hat{\phi}_{\vec{k},\sigma} = 0 \tag{130}$$

$$\frac{d^2 \hat{\phi}_{\vec{k},\sigma}}{d\upsilon^2} - \frac{1}{a}\frac{d^2 a}{d\upsilon^2} \hat{\phi}_{\vec{k},\sigma} = 0 \tag{131}$$

One can see that they are invariant with respect to Wick rotation $\eta = i\upsilon$ (96). In other words, the superhorizon waves do not feel the existence of topologically impenetrable barrier between Lorentzian and Euclidean spaces, which means that this barrier becomes permeable to these waves. Therefore, the boundary conditions (25) for classical waves and (107) for gravitons allow us to make an analytical continuation from Euclidean space to Lorentzian space. These boundary conditions must be satisfied at the barrier $k = 0$. This analytical continuation is possible also because of invariance of the equation of state (113) with respect to Wick rotation. Therefore, the analytic continuation of the Euclidean solution to Lorentzian space of real time gives us the accelerated de Sitter expansion in real time (110). In other words, the energy needed to accelerate the expansion in real time comes from Euclidean space, i.e., from "nothing".

The following scenario of cosmological evolution was proposed in our work [2]. "A flat inflationary Universe could have been formed by tunneling from "nothing". After that, it should evolve in accordance to inflation scenarios that are beyond the scope of this paper. Then the standard Big Bang cosmology starts and lasts as long as the Universe begins to become empty again. As the Universe ages and is emptied, the same mechanism of tunneling that gave rise to the empty Universe at the beginning, gives now birth to dark energy. This mechanism is switched on after the energy density of matter has dropped below a critical level. After that, to the extent that the space continues to empty, expansion takes place faster and faster, and gradually it becomes again exponentially fast (de Sitter)". Since the analytic continuation is valid both for a passage from Euclidean space to Lorentzian space and vice versa (this is, so to speak, a "two way street") it provides a possibility for the empty Universe (that has completed its cosmological evolution) to be able to tunnel back to "nothing". After that, the entire scenario can be repeated indefinitely.

**Acknowledgments:** I would like to express my appreciation to Walter Sadowski for invaluable help in the preparation of the manuscript. I am grateful to Daniel Usikov and Boris Vayner for useful discussions. I would like to thank all three anonymous referees for valuable comments that helped improve this work.

**Conflicts of Interest:** The authors declare no conflict of interest.

**Appendix. Stochastic Nonlinear Gravitational Waves in an Isotropic Universe**[15]

**A. Definitions of the Background and Fluctuations**

We consider the situation when the geometric characteristics of the space-time metric $\hat{g}_{ik}$, the connectivity $\hat{\Gamma}^l_{ik}$, and the curvature $\hat{R}_{ik}$ -are fluctuating functions for some physical reasons. At the same time, we assume that there exist regular regularly determined components of these functions $g_{ik}$, $\Gamma^l_{ik}$, $R_{ik}$. In both cases we assume that the standard relations of Riemannian geometry are satisfied:

$$\hat{\Gamma}^l_{ik} = \frac{1}{2}\hat{g}^{lm}(\hat{g}_{mi,k} + \hat{g}_{mk,i} - \hat{g}_{ik,m})$$
$$\hat{R}_{ik} = \hat{\Gamma}^l_{ik,l} - \hat{\Gamma}^l_{il,k} + \hat{\Gamma}^l_{ik}\hat{\Gamma}^m_{lm} - \hat{\Gamma}^l_{im}\hat{\Gamma}^m_{kl} \tag{A1}$$

$$\Gamma^l_{ik} = \frac{1}{2}g^{lm}(g_{mi,k} + g_{mk,i} - g_{ik,m})$$
$$R_{ik} = \Gamma^l_{ik,l} - \Gamma^l_{il,k} + \Gamma^l_{ik}\Gamma^m_{lm} - \Gamma^l_{im}\Gamma^m_{kl} \tag{A2}$$

Because of the nonlinearity of the Einstein equations, the question of how to extract the background geometry from the fluctuating geometry is not trivial. A mathematically unambiguous procedure for isolating deterministic macroscopic geometry from a complete fluctuating geometry is offered by the functional integration method. Using this method, one can rigorously prove that the fluctuating component is distinguished multiplicatively by the densities of contravariant and covariant metrics ([3], Section II.C).

$$\sqrt{-\hat{g}}\,\hat{g}^{ik} = \sqrt{-g}\,g^{il}X^k_l \qquad \frac{\hat{g}^{ik}}{\sqrt{-\hat{g}}} = \frac{g^{il}}{\sqrt{-g}}Y^l_k \tag{A3}$$

$$X^k_l \equiv (\exp\psi)^k_l = \delta^k_l + \hat{\psi}^k_l + \frac{1}{2!}\hat{\psi}^m_l\hat{\psi}^k_m + \frac{1}{3!}\hat{\psi}^m_l\hat{\psi}^n_m\hat{\psi}^k_n + ...$$
$$\hat{Y}^l_k \equiv (\exp(-\psi))^l_k = \delta^l_k - \hat{\psi}^l_k + \frac{1}{2!}\hat{\psi}^l_m\hat{\psi}^m_k - \frac{1}{3!}\hat{\psi}^l_m\hat{\psi}^m_n\hat{\psi}^n_k + ... \tag{A4}$$

where $X^k_l$ and $Y^l_k$ are mutually inverse matrix exponents connected by the following identity

$$X^l_i Y^k_l \equiv \delta^k_i \tag{A5}$$

By definition, the mean value of the random function $\psi^k_i$ in the statistical ensemble is zero

$$<\psi^k_i> = 0 \tag{A6}$$

Simultaneously with the definition of background and fluctuating dynamic variables, the functional integral method fixes the form of the original complete Einstein equations

---

[15] The material constituting the main content of the present Appendix was prepared by the late Grigory Vereshkov. It was to be included in our joint paper [3]. We did not include it in the final version of [3], since [3] was entirely devoted to quantum effects.

$$\sqrt{-\hat{g}}\,\hat{g}^{il}\hat{R}_{lk} - \frac{1}{2}\delta_i^k\sqrt{-\hat{g}}\,\hat{g}^{ml}\hat{R}_{lm} = 0 \tag{A7}$$

In a theory with a deterministic metric, equations (A7) can be subjected to transformations that have the status of the identities (e.g., lowering and rising of the metric indices, reducing by the common factor $\sqrt{-\hat{g}}$). The situation is different in the statistical theory. Any transformations of non-linear equations consisting of their multiplication or division by a function containing a random component are forbidden because after averaging these actions lead to different equations.

The self-consistent set of equations for the background and fluctuations can be obtained from (A7) by a standard method. The equations for background read

$$<\sqrt{-\hat{g}}\,\hat{g}^{il}\hat{R}_{lk}> - \frac{1}{2}\delta_i^k <\sqrt{-\hat{g}}\,\hat{g}^{ml}\hat{R}_{lm}> = 0 \tag{A8}$$

The equations for fluctuations read

$$\sqrt{-\hat{g}}\,\hat{g}^{il}\hat{R}_{lk} - <\sqrt{-\hat{g}}\,\hat{g}^{il}\hat{R}_{lk}> - \frac{1}{2}\delta_i^k\sqrt{-\hat{g}}\,\hat{g}^{ml}\hat{R}_{lm} + \frac{1}{2}\delta_i^k <\sqrt{-\hat{g}}\,\hat{g}^{ml}\hat{R}_{lm}> = 0 \tag{A9}$$

We emphasize that the self-consistency of Equations (A8) and (A9), i.e., the possibility of their joint solution for finding the functions $g_{ik}$ and $\psi_i^k$) can be obtained when three following conditions are fulfilled. (1) Parameterization of the metric tensor that divides the deterministic and random components is carried out according to the multiplicative rule (A3) and (A4); (2) The initial equations are written in the form (A7); (3) All averaging carried out taking into account the definition (A6).

Absence of any of these conditions leads to a mathematically contradictory system of equations that has no solution. For example, Equations (A8) and (A9) are contradictory if we substitute in them the linearly parameterized metric $\hat{g}_{ik} = g_{ik} + h_{ik}$ ; $<h_{ik}> = 0$ Note that the method of functional integration easily reveals the mathematical nature of this contradiction. A self-consistent description of the background and fluctuations assumes the factorization of the measure of the functional integral. If so, it is possible to sequentially integrate over the fluctuating and background fields. Integration over the fluctuations corresponds to solving the equations for fluctuations and substituting the resulting solutions into the averaged background equations. Integration over the background corresponds to the solution of the background equations taking into account the backreaction of fluctuations on the background. If the measure of the functional integral is not factorized (and this takes place with the parameterization $\hat{g}_{ik} = g_{ik} + h_{ik}$), then it is impossible to perform the described operations, i.e., to get the joint solution of the equations for fluctuations and the background. Neglecting this fact and forcibly putting an incorrect parameterization into Einstein's equations, we get a mathematical contradiction: Bianchi's identities for the background geometry are not satisfied in the equations of motion for fluctuations. As was already mentioned in the Introduction, the authors of all papers (with no exceptions) on the backreaction of fluctuations on the background published during the time of 1977–2008 used the linear parameterization.

Substitution of parameterization (A4) into equations (A7) reads

$$\sqrt{\frac{\hat{g}}{g}}\,\hat{g}^{kl}\hat{R}_{il} \equiv \left\{\hat{X}_l^{\ k}R_i^{\ l} + \frac{1}{2}\left[\hat{Y}^m_{\ i}\left(\hat{X}_n^{\ l}\hat{X}_m^{\ k;m} - \hat{X}_n^{\ k}\hat{X}_m^{\ l;n}\right) - \frac{1}{2}\hat{Y}^q_{\ p}\hat{X}_m^{\ l}\hat{X}_q^{\ p;m} - X^{kl}_{\ ;i}\right]_{;l} - \right.$$
$$\left. - \frac{1}{4}\left(\hat{Y}^p_{\ n}\hat{Y}^q_{\ m} - \frac{1}{2}\hat{Y}^p_{\ m}\hat{Y}^q_{\ n}\right)\hat{X}_l^{\ k}\hat{X}_{p;i}^{\ m}\hat{X}_q^{\ n;l} + \frac{1}{2}\hat{Y}^q_{\ p}\hat{X}_{i\ ;m}^{\ p}\hat{X}_q^{\ m;k}\right\} = 0 \tag{A10}$$

There is also the exact solution to Equation (A5) which reads

$$Y_i^k = \frac{1}{\Delta}[\frac{1}{6}\delta_i^k(X^3 - 3XX_l^m X_m^l + 2X_l^m X_m^n X_n^l)$$
$$-\frac{1}{2}X_i^k(X^2 - X_m^l X_l^m) + XX_i^l X_l^k - X_i^l X_l^m X_m^k]$$
(A11)

$$\Delta = \frac{1}{24}[X^4 - 6X^2 X_l^m X_m^l + 8XX_l^m X_m^n X_n^l + 3(X_l^m X_m^l)^2 - 6X_l^m X_m^n X_n^s X_s^l] \equiv e^\psi$$

In the deterministic theory (with no use of fluctuations), Equation (A10) are valid also. In this case, instead of the term "background metric", it is more correct to use the term "given basic metric." In fact, in Equations (A10) and (A11) we are dealing with the parameterization of the so-called two-metric theory, which in its time was discussed in detail in the literature. In a deterministic theory, the basic metric is fixed and then used for operations with indices ($X_l^k$ and $Y_i^k$ are tensors in the basis space) and covariant differentiation. Taking into account Equation (A11), Equation (A10) represent a system of equations for the tensor $X_l^k$, written in terms of the same tensor. For this tensor, without loss of generality, we can specify a gauge that is covariant in the basis space. In fact, this is Fock's gage.

$$X_{i;k}^k = 0 \tag{A12}$$

The choice of the basis space fixes the class of topologies of Riemannian spaces studied in the framework of Einstein's Equation (A10). There is nothing new in comparison with the traditional formulation of Einstein's theory. The only difference is that in the traditional formulation the choice of topology is fixed by restrictions on the metric in the process of solving Einstein's equations, and in the formulation (A10)—the same is done before solving the equations by fixing the basic metric. In those cases where the Einstein equations can really be solved, the choice of algorithm is a matter of convenience.

In the theory dealing with the concept of stochastic geometry, Equation (A10) have the status of formal relations, demonstrating the fundamental possibility of constructing a statistical theory that does not use the hypothesis of smallness of fluctuations. However, the theory acquires concrete content only after carrying out procedures (A8) and (A9), which presuppose a mathematically unambiguous definition of the averaging operation. Unfortunately, problems arise with the latter, since the fluctuating variable is "hidden" in the matrix function $X_l^k$, and can only be separated out in explicit form (and carrying out averaging in explicit form) after the expansion of the matrix function in a series in its tensor argument.

We represent Equation (A10) in the form in which the derivative of the fluctuating variable with respect to the coordinates is in an explicit form. When this transformation is performed, some technical problems arise, solutions of which are clarified within the framework of the functional integration formalism. The problems come from the fact that we are dealing with the complex object

$$X_i^k = X_i^k[\psi_l^m(x^n)]$$

where $\psi_l^m$ is the tensor argument of a matrix function and $x^n$ is the argument of argument or the vector parameter of matrix function. As soon as we try to work with the tensor argument, we must immediately take into account the obvious fact: the differentiation of the matrix function with respect to the tensor argument and the differentiation of the tensor argument with respect to the vector parameter generates a wide set of non-commuting matrices. In particular, the operations of expanding a matrix function in a series and differentiating a matrix function with respect to a vector argument are non-commuting. Therefore, it is necessary to fix the correct sequence of these operations. The method of functional integration allows us to formulate these rules. The first two rules are the so-called rules of truncated tensors

(1) In Einstein's Equation (A10), the sorting of free indices is carried out according to the rule $i \leq k$
(2) The objects of the theory are mixed tensors $X_i^{\ k}$ and $\psi_i^{\ k}$, in which the sorting of free indices is carried out according to the same rule $i \leq k$. These rules provide the same number of independent equations and independent functions as in the standard formulation of Einstein's equations.
(3) The derivative of the matrix function with respect to the vector parameter is determined before the expansion of the matrix function with respect to the tensor argument, i.e.,

$$X_{i;l}^{\ k} = X_i^{\ m}\psi_{m;l}^{\ k} \tag{A13}$$

The expansions of matrix functions in a series with respect to the tensor variable are carried out only after all the differential operations are performed. All the rules (and we emphasize—the most important rule (A13)) were verified on exact solutions of Einstein's equations.

Taking into account (A13), Equation (A10) read

$$\sqrt{\frac{\hat{g}}{g}}\hat{g}^{kl}\hat{R}_{il} \equiv \left\{ X_l^{\ k}R_i^{\ l} + \frac{1}{2}(X_n^{\ l}\psi_i^{\ k;n} - X_n^{\ k}\psi_i^{\ l;n} - \frac{1}{2}\delta_i^k X_m^{\ l}\psi^{;m})_{;l} - \frac{1}{2}(X_l^{\ m}\psi_{m;i}^{\ k})^{;l} \right.$$
$$\left. - \frac{1}{4}[X_l^{\ k}\psi_{m;i}^{\ n}\psi_n^{\ m;l} - \frac{1}{2}X_l^{\ k}\psi_{;i}\psi^{;l} - 2X_l^{\ n}\psi_{n;m}^{\ l}\psi_i^{\ m;k}\right\} = 0 \tag{A14}$$

**B. Stochastic Nonlinear Gravitational Waves over the FLRW Background**

We take the FLRW metric of the flat Universe as the background metric. In this case, we have

$$ds^2 = dt^2 - a(t)^2\gamma_{\alpha\beta}dx^\alpha dx^\beta = dt^2 - a(t)^2(dx^2 + dy^2 + dz^2)$$
$$R_0^{\ 0} = -3\frac{\ddot{a}}{a} \qquad R_\alpha^{\ \beta} = -\delta_\alpha^{\ \beta}(\frac{\ddot{a}}{a} + 2\frac{\dot{a}^2}{a^2}) \tag{A15}$$

In the synchronous gage we get

$$\psi_0^{\ 0} = 0 \qquad \psi_0^{\ \alpha} = 0 \tag{A16}$$

Einstein's equations in the explicit form before averaging read

$$\sqrt{\frac{\hat{g}}{g}}\hat{g}^{0l}\hat{R}_{0l} \equiv -3\frac{\ddot{a}}{a} - \frac{1}{4}\{\ddot{\psi} - \frac{\dot{a}}{a}\dot{\psi} - \frac{1}{a^2}(X_\mu^{\ \nu}\psi^{,\mu})_{,\nu}\} - \frac{1}{4}\dot{\psi}_\mu^{\ \nu}\dot{\psi}_\nu^{\ \mu} + \frac{1}{8}\dot{\psi}^2 = 0 \tag{A17}$$

$$\sqrt{\frac{\hat{g}}{g}}\hat{g}^{\beta l}\hat{R}_{\alpha l} \equiv -\delta_\alpha^{\ \beta}(\frac{\ddot{a}}{a} + 2\frac{\dot{a}^2}{a^2}) + \frac{1}{2}\{-\frac{1}{2}\delta_\alpha^{\ \beta}[\ddot{\psi} + 3\frac{\dot{a}}{a}\dot{\psi} - \frac{1}{a^2}(X_\mu^{\ \nu}\psi^{,\mu})_{,\nu}] + \ddot{\psi}_\alpha^{\ \beta} + 3\frac{\dot{a}}{a}\dot{\psi}_\alpha^{\ \beta}$$
$$- \frac{1}{a^2}(X_\mu^{\ \nu}\dot{\psi}_\alpha^{\ \beta,\mu} - X_\mu^{\ \beta}\psi_\alpha^{\ \nu,\mu} - X_\mu^{\ \nu}\psi_\alpha^{\ \mu,\beta})_{,\nu}\} \tag{A18}$$
$$+ \frac{1}{4a^2}[X_\lambda^{\ \beta}\psi_{\mu,\alpha}^{\ \nu}\psi_\nu^{\ \mu,\lambda} - \frac{1}{2}X_\lambda^{\ \beta}\psi_{,\alpha}\psi^{,\lambda} - 2X_\alpha^{\ \nu}\psi_{\nu,\mu}^{\ \lambda}\psi^{\mu,\beta}_{\ \ \lambda}]$$

$$\sqrt{\frac{\hat{g}}{g}}\hat{g}^{0l}\hat{R}_{\alpha l} \equiv \frac{1}{2}(-\dot{\psi}_{\alpha,\mu}^{\ \mu} + \frac{\dot{a}}{a}\psi_{,\alpha}) - \frac{1}{4}(\psi_{\nu,\alpha}^{\ \mu}\dot{\psi}_\mu^{\ \nu} - \frac{1}{2}\psi_{,\alpha}\dot{\psi} - 2X_\alpha^{\ \lambda}\psi_{\lambda,\mu}^{\ \nu}\dot{\psi}_\nu^{\ \mu}) \tag{A19}$$

In Equations (A17)–(A19) dots are derivatives over time. All operations with the spatial indexes are conducting with Euclidian metric. Averaging of Equations (A17) and (A18) leads to the fact that all terms in the curly brackets are zeroed, and we get

$$-3\frac{\ddot{a}}{a} = \frac{1}{4} <\dot{\psi}_\mu{}^\nu \dot{\psi}_\nu{}^\mu - \frac{1}{2}\dot{\psi}^2> \tag{A20}$$

$$3\frac{\ddot{a}}{a} + 6\frac{\dot{a}^2}{a^2} = \frac{1}{4a^2} < X_\lambda{}^\alpha \psi_{\mu,\alpha}{}^\nu \psi_\nu{}^{\mu,\lambda} - \frac{1}{2} X_\lambda{}^\alpha \psi_{,\alpha} \psi^{,\lambda} - 2 X_\alpha{}^\nu \psi_{\nu,\mu}{}^\lambda \psi_\lambda{}^{\mu,\alpha} > \tag{A21}$$

We can define the energy density and pressure of nonlinear gravitational wave medium as follows

$$3\frac{\dot{a}^2}{a^2} = \kappa \rho_g \tag{A22}$$

$$\frac{\ddot{a}}{a} = -\frac{\kappa}{12}(\rho_g + 3 p_g) \tag{A23}$$

where $\kappa = 8\pi G$, speed of light $c = 1$. Thus, the energy density of such a nonlinear gravitational wave medium reads

$$\kappa \rho_g = \frac{1}{8} < \dot{\psi}_\mu{}^\nu \dot{\psi}_\nu{}^\mu - \frac{1}{2}\dot{\psi}^2 + \frac{1}{a^2}(X_\mu{}^\nu \psi_\lambda{}^{\sigma,\mu} \psi_{\sigma,\nu}{}^\lambda - 2 X_\mu{}^\nu \psi_\lambda{}^{\sigma,\mu} \psi_{\nu,\sigma}{}^\lambda - \frac{1}{2} X_\mu{}^\nu \psi_{,\nu} \psi^{,\mu}) > \tag{A24}$$

The pressure of such a nonlinear gravitational wave medium can be found from Equations (A20), (A23) and (A24). Finally, we turn to the equations for the fluctuations. They need to be divided into equations of constraints and equations of proper dynamics. The constraint equations can be obtained from (A17)–(A21). We do not present them here. The equations of proper dynamics follow from (A18). They read

$$\ddot{\psi}_\alpha{}^\beta + 3\frac{\dot{a}}{a}\dot{\psi}_\alpha{}^\beta - \frac{1}{a^2}(X_\nu{}^\mu \psi_\alpha{}^{\beta,\mu} - X_\mu{}^\beta \psi_\alpha{}^{\nu,\mu} X_\nu{}^\nu \psi_\alpha{}^{\mu,\beta})_{,\nu} - \frac{1}{2}\delta_\alpha{}^\beta [\ddot{\psi} + 3\frac{\dot{a}}{a}\dot{\psi} - \frac{1}{a^2}(X_\mu{}^\nu \psi^{,\mu})_{,\nu} = \\ -\frac{1}{2a^2}[X_\lambda{}^\beta \psi^\nu{}_{\mu,\alpha} \psi_\nu{}^{\mu,\lambda} - \frac{1}{2} X_\lambda{}^\beta \psi_{,\alpha} \psi^{,\lambda} - 2 X_\alpha{}^\nu \psi_{\nu,\mu}{}^\lambda \psi_\lambda{}^{\mu,\beta}] + \frac{1}{2a^2} < [X_\lambda{}^\beta \psi^\nu{}_{\mu,\alpha} \psi_\nu{}^{\mu,\lambda} - \frac{1}{2} X_\lambda{}^\beta \psi_{,\alpha} \psi^{,\lambda} - 2 X_\alpha{}^\nu \psi_{\nu,\mu}{}^\lambda \psi_\lambda{}^{\mu,\beta}] > \tag{A25}$$

Equations (A22), (A24) and (A25) form a self-consistent set of equations describing a medium consisting of nonlinear gravitational waves in the space-time with the FLRW metric.

As is known, non-polynomial field theories in which it is necessary to perform statistical averaging include a wide class of models of quantum field theory that are used to describe strong interactions at the level of compound particles-hadrons. Statistical averaging in this case is averaging over the quantum ensemble. No other methods have been proposed in this area of theoretical physics, except averaging after the expansion of non-polynomial functions in a series. In particular, this circumstance hampers progress in the study of strong interactions. Similar problems exist in string theory. It should be recognized that in the absence of a regular method of investigating the equations of a theory that does not use the operation of expanding a non-polynomial function in a series, the very notion of a strong fluctuation is controversial. We have only one argument: the problem of averaging in non-polynomial theories is typical. As in all other cases, we will try to move forward using more or less plausible assumptions about the properties of the solution of the equations of this theory.

## C. Stochastic Gravitational Waves over the FLRW Background in the Quasi-Linear Approximation

In this approximation, the dynamics of gravitational waves are described by Equation (A25) in the linear approximation but the backreaction of gravitational waves on the background metric is described by Equations (A22)–(A24) in which the second order terms are retained. This means that the backreaction of waves affect the background metric which in turn affect the state of gravitational waves. This also means that the interaction of gravitational waves is taken into account only through self-consistent background gravitational field. Assuming that gravitational waves are gauged and contain only two degree of freedom, we get

$$\psi_0^{\ 0} = \psi_\alpha^{\ 0} = 0; \quad \psi \equiv \psi_\alpha^{\ \alpha} = 0; \quad \psi_{\alpha,\beta}^{\ \beta} = 0 \tag{A26}$$

Taking into account (A26), Equations (A20), (A21) and (A25) read

$$\ddot{\psi}_\alpha^{\ \beta} + 3\frac{\dot{a}}{a}\dot{\psi}_\alpha^{\ \beta} - \frac{1}{a^2}\psi_{\alpha,\gamma}^{\ \beta,\gamma} = 0 \tag{A27}$$

$$\frac{\ddot{a}}{a} + 2\frac{\dot{a}^2}{a^2} = \frac{1}{12a^2} <\psi^\beta_{\ \alpha,\gamma}\psi_\beta^{\ \alpha,\gamma}> \tag{A28}$$

$$3\frac{\dot{a}^2}{a^2} = \frac{1}{8}<\dot{\psi}_\alpha^{\ \beta}\dot{\psi}_\beta^{\ \alpha} + \frac{1}{a^2}\psi_{\alpha,\gamma}^{\ \beta}\psi_\beta^{\ \alpha,\gamma}> \tag{A29}$$

From (A27)–(A29) we get the energy density and pressure of the gravitational wave medium that read

$$\kappa\rho_g = \frac{1}{8}<\dot{\psi}_\alpha^{\ \beta}\dot{\psi}_\beta^{\ \alpha} + \frac{1}{a^2}\psi_{\alpha,\gamma}^{\ \beta}\psi_\beta^{\ \alpha,\gamma}> \tag{A30}$$

$$\kappa p_g = \frac{1}{8}<\dot{\psi}_\alpha^{\ \beta}\dot{\psi}_\beta^{\ \alpha} - \frac{1}{3a^2}\psi_{\alpha,\gamma}^{\ \beta}\psi_\beta^{\ \alpha,\gamma}> \tag{A31}$$

Equations (A27) and (A28) or Equations (A27) and (A29) form the self-consistent system of equations describing the backreaction of classical gravitational waves on FLRW background and the behavior of these waves on the background affected by them. In other words, in our self-consistent approach the state of gravitational waves is determined by their interaction with the background geometry, and the background geometry, in turn, depends on the state of CGW.